\documentclass[12pt,a4paper]{article}

\usepackage[margin=25mm]{geometry}
\usepackage[font=sf]{caption} 

\usepackage{amsmath}
\usepackage{amsfonts}
\usepackage{amssymb}
\usepackage{siunitx}
\usepackage{verbatim}
\usepackage{hyperref} 
\usepackage[
backend=biber,
style=nature,
]{biblatex}
\usepackage{tikz}
\usepackage{filecontents}
\usepackage{amsthm}

\usepackage{float}
\usepackage{graphics}
\usepackage{graphicx}
\usepackage[skip=0pt]{caption}
 \usepackage{algorithm} 
\usepackage[noend]{algorithmic}
\renewcommand{\algorithmicrequire}{\textbf{Input:}}
\renewcommand{\algorithmicensure}{\textbf{Output:}}
\usepackage{bm}
\usepackage{color}
\usepackage{multirow}
\usepackage{makecell}
\usepackage{balance}
\usepackage{pdfx}
\usepackage{subcaption}
\usepackage{autobreak}

\usepackage{booktabs}
\usepackage{graphbox}

\newcommand{\myparatight}[1]{\smallskip\noindent{\bf {#1}:}~}
\title{\Large \bf Stable Signature is Unstable: Removing Image Watermark from Diffusion Models}
\author{
  Yuepeng Hu, Zhengyuan Jiang, Moyang Guo, Neil Gong \\
  Duke University \\
  \{yuepeng.hu, zhengyuan.jiang, moyang.guo, neil.gong\}@duke.edu
  }
\date{}

\begin{document}
\maketitle

\begin{abstract}
Watermark has been widely deployed by industry to detect AI-generated images. A recent watermarking framework called \emph{Stable Signature} (proposed by Meta) roots watermark into the parameters of a diffusion model's decoder such that its generated images are inherently watermarked. Stable Signature makes it possible to watermark images generated by \emph{open-source} diffusion models and was claimed to be robust against removal attacks. In this work, we propose a new attack to remove the watermark from a diffusion model by fine-tuning it. Our results show that our attack can effectively remove the watermark from a diffusion model such that its generated images are non-watermarked, while maintaining the visual quality of the generated images. Our results highlight that Stable Signature is not as stable as previously thought. 
\end{abstract}

\section{Introduction}
With the rapid development of generative AI (GenAI), it becomes increasingly more difficult to distinguish AI-generated and non-AI-generated images. The misuse of AI-generated images presents a significant risk of spreading misinformation. Watermarking~\cite{bi2007robust,zhu2018hidden,UDH,2019stegastamp,fernandez2023stable,wen2023tree} has emerged as a crucial technology for detecting AI-generated images and been widely deployed by industry. For instance, OpenAI incorporates a watermark into images generated by DALL-E~\cite{ramesh2021zero}; Stability AI deploys a watermarking technique in Stable Diffusion~\cite{stable-diffusion}; and Google has introduced SynthID as a watermarking solution for images generated by Imagen~\cite{saharia2022photorealistic}. In watermark-based detection,  a watermark is embedded in  AI-generated images before they are accessed by users. During  detection, if the same watermark can be extracted from an image, it is identified as AI-generated.

Image watermark can be categorized into three groups based on the timing when watermark is embedded into AI-generated images: \emph{post-generation}, \emph{pre-generation}, and \emph{in-generation}. Post-generation watermark~\cite{luo2020distortion,bi2007robust,zhu2018hidden,UDH,al2007combined,2019stegastamp} embeds a watermark into an image after the image has been generated, while pre-generation watermark~\cite{wen2023tree} embeds a watermark into the initial noisy latent vector of a diffusion model. However, these watermarking methods are vulnerable when  the diffusion models are open-source. In particular, an attacker can easily remove the watermarking components from the open-source diffusion model to generate non-watermarked images. In contrast, in-generation watermark (e.g., Stable Signature~\cite{fernandez2023stable}) roots watermark directly into the parameters of a diffusion model's decoder. It enables the images generated by the diffusion model to be inherently watermarked without introducing any external watermarking components. This method is particularly suited for watermarking images generated by open-source diffusion models.

Watermark removal attacks aim to remove watermarks from watermarked images, and can be divided into two main types: \emph{per-image-based} and \emph{model-targeted}. Per-image-based attacks~\cite{jiang2023evading, an2024benchmarking,lukas2023leveraging,zhao2023invisible,saberi2023robustness} add a carefully crafted perturbation to each watermarked image individually. These removal attacks need to process watermarked images one by one, which is highly inefficient when removing watermarks from a large volume of watermarked images. In contrast, model-targeted  attacks directly modify a diffusion model's parameters to make its generated images  non-watermarked. For instance, the authors of Stable Signature~\cite{fernandez2023stable} also proposed a model-targeted removal attack, called \emph{model purification (MP)}, to attack Stable Signature. However, MP requires access to the diffusion model's encoder and is not applicable when the encoder is not open-source. Moreover,  MP significantly deteriorates image quality~\cite{fernandez2023stable}, based on which Stable Signature was claimed to be robust against model-targeted removal attacks.

\begin{figure}[t!]
\centering
\begin{subfigure}{0.19\linewidth}
  \centering
  \subfloat[Clean]{\includegraphics[width=\linewidth]{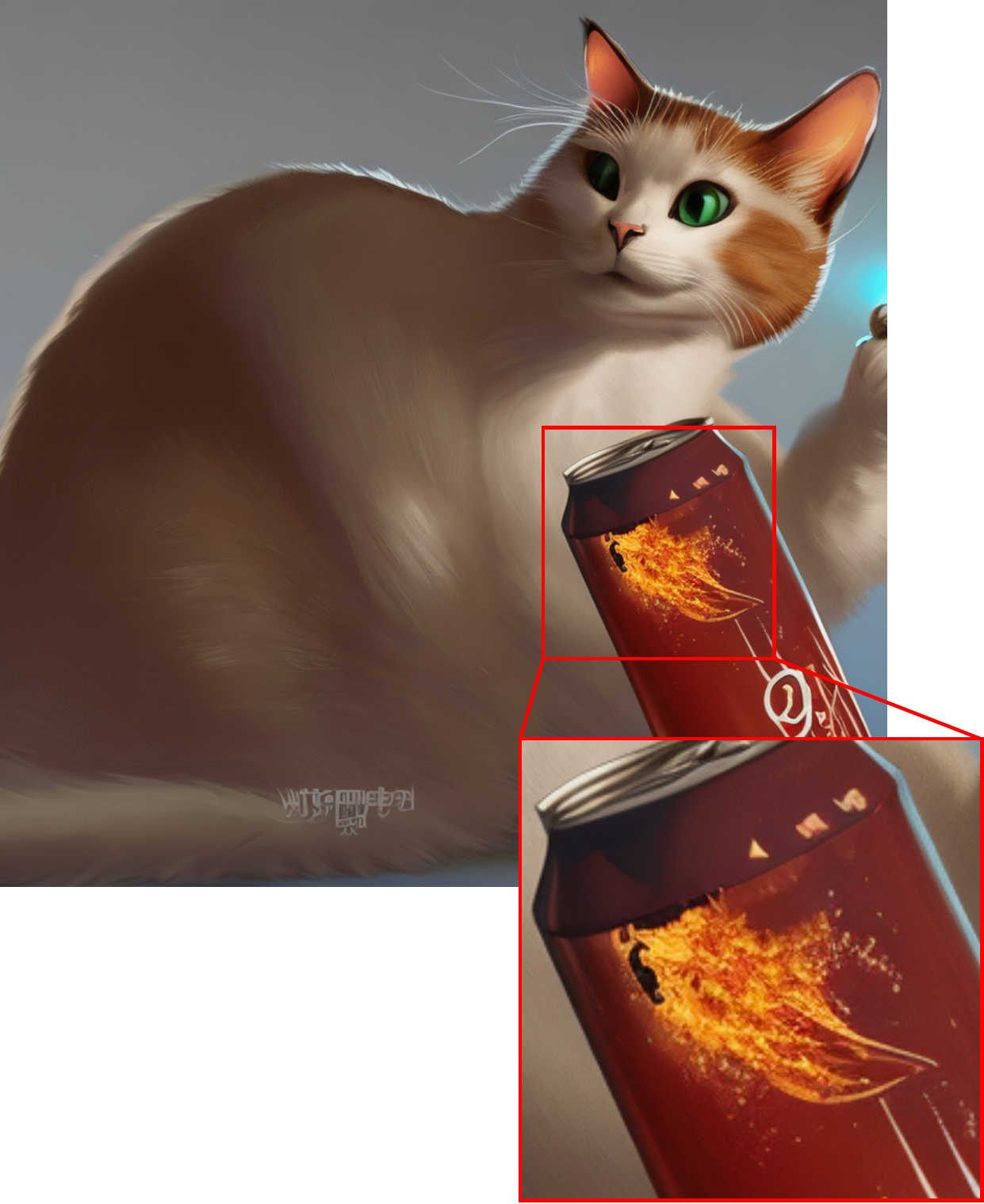}}
\end{subfigure}
\centering
\begin{subfigure}{0.19\linewidth}
  \centering
  \subfloat[Watermarked]{\includegraphics[width=\linewidth]{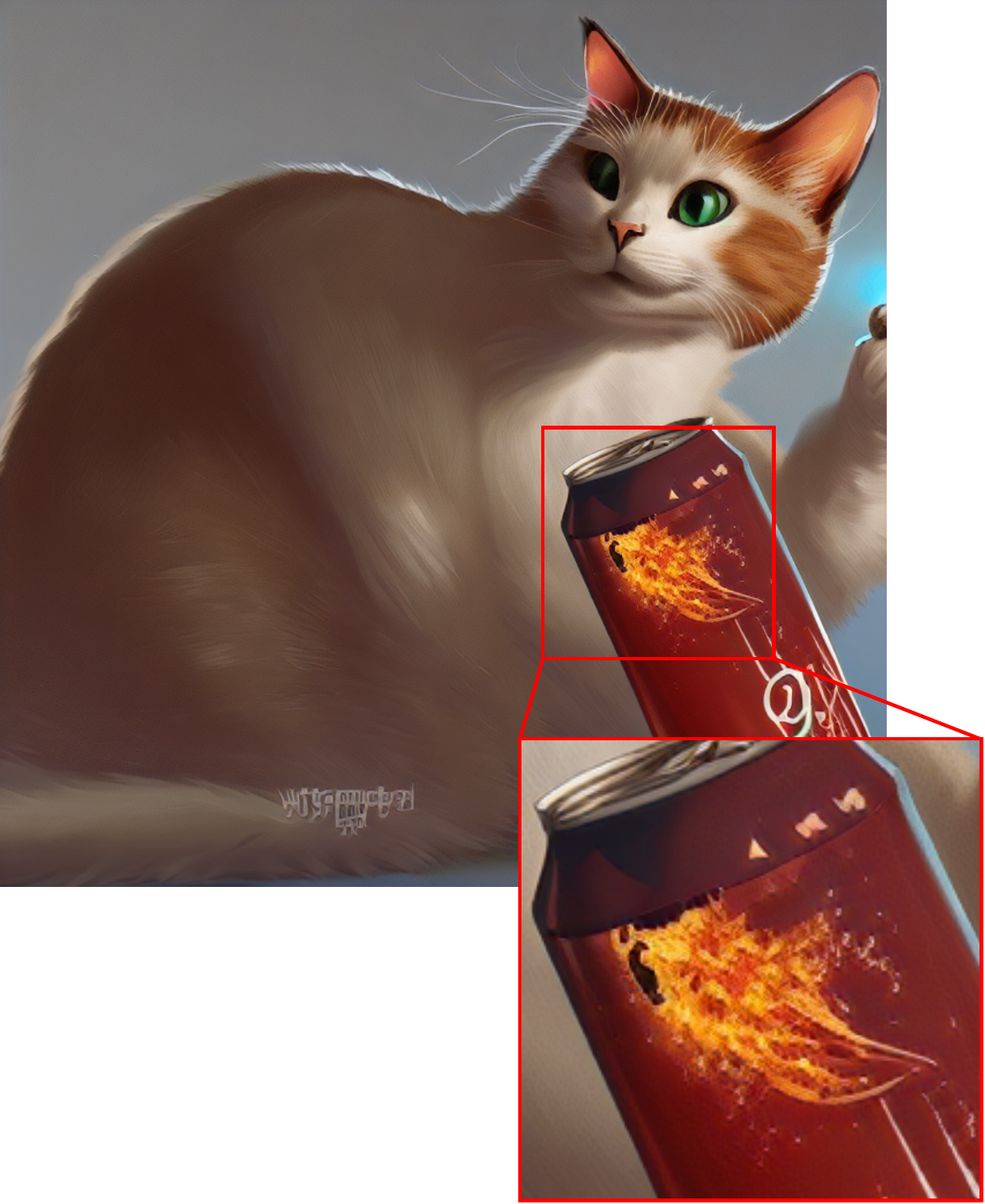}}
\end{subfigure}
\begin{subfigure}{0.19\linewidth}
  \centering
  \subfloat[MP]{\includegraphics[width=\linewidth]{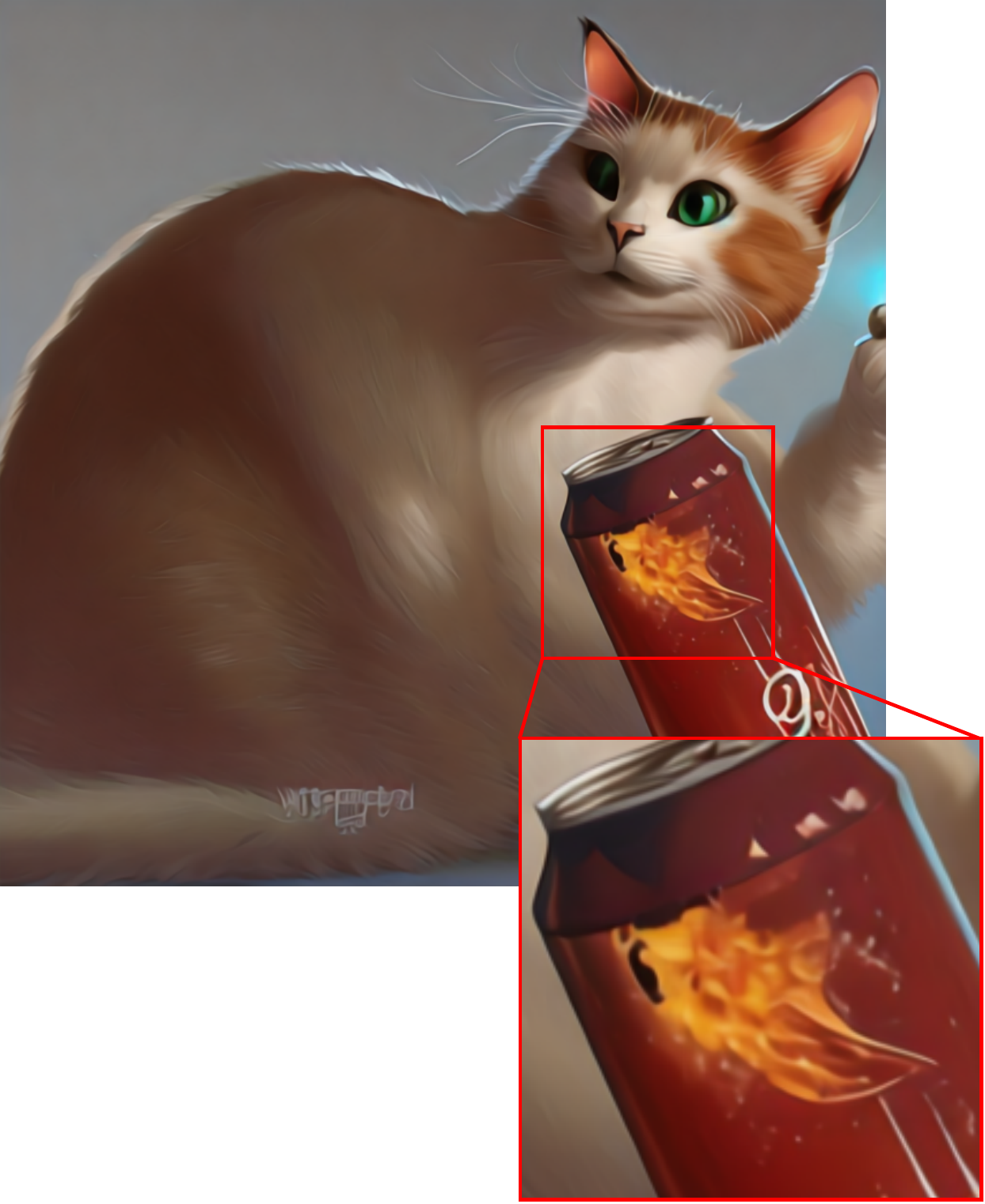}}
\end{subfigure}
\begin{subfigure}{0.19\linewidth}
  \centering
  \subfloat[E-aware]{\includegraphics[width=\linewidth]{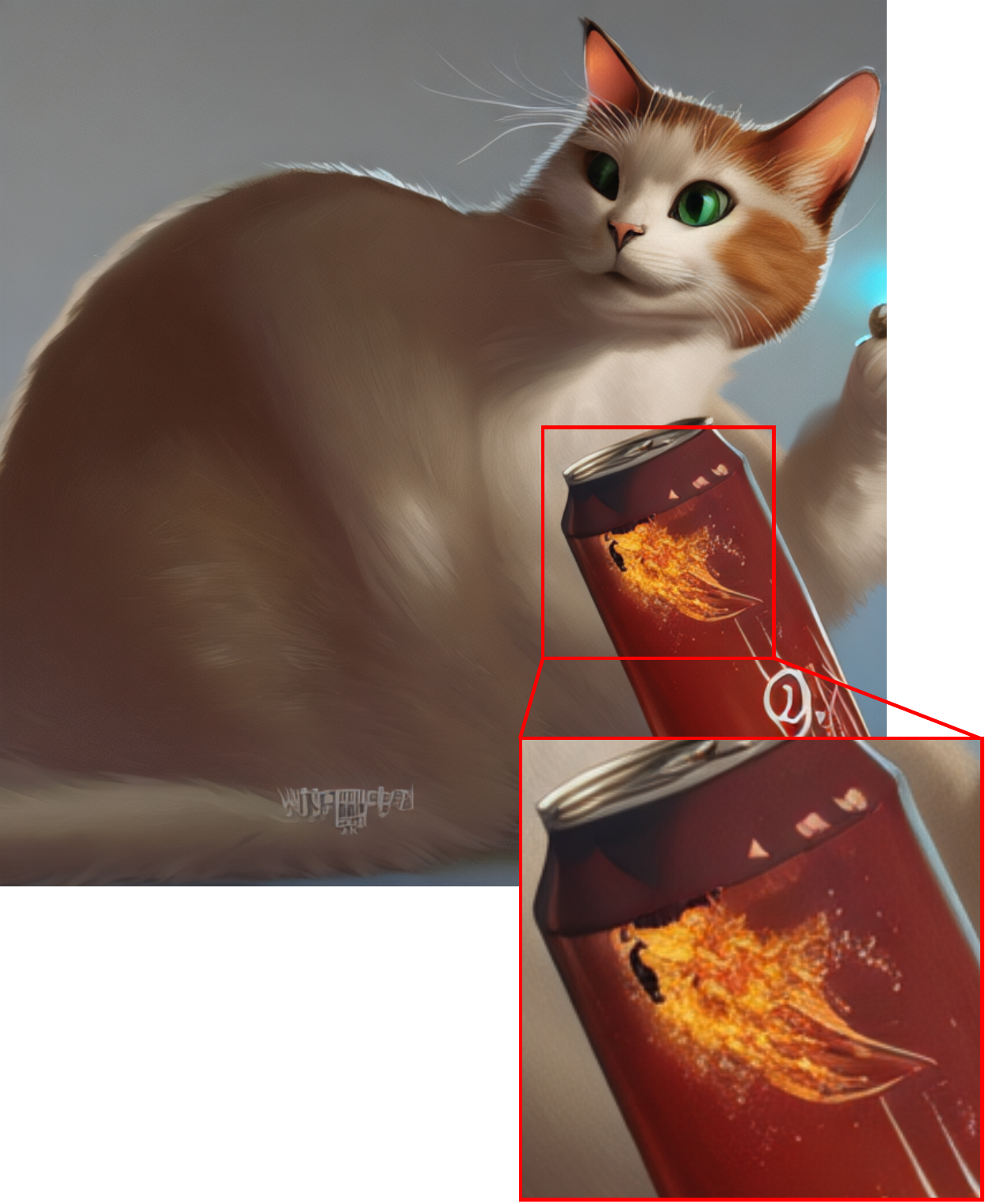}}
\end{subfigure}
\begin{subfigure}{0.19\linewidth}
  \centering
  \subfloat[E-agnostic]{\includegraphics[width=\linewidth]{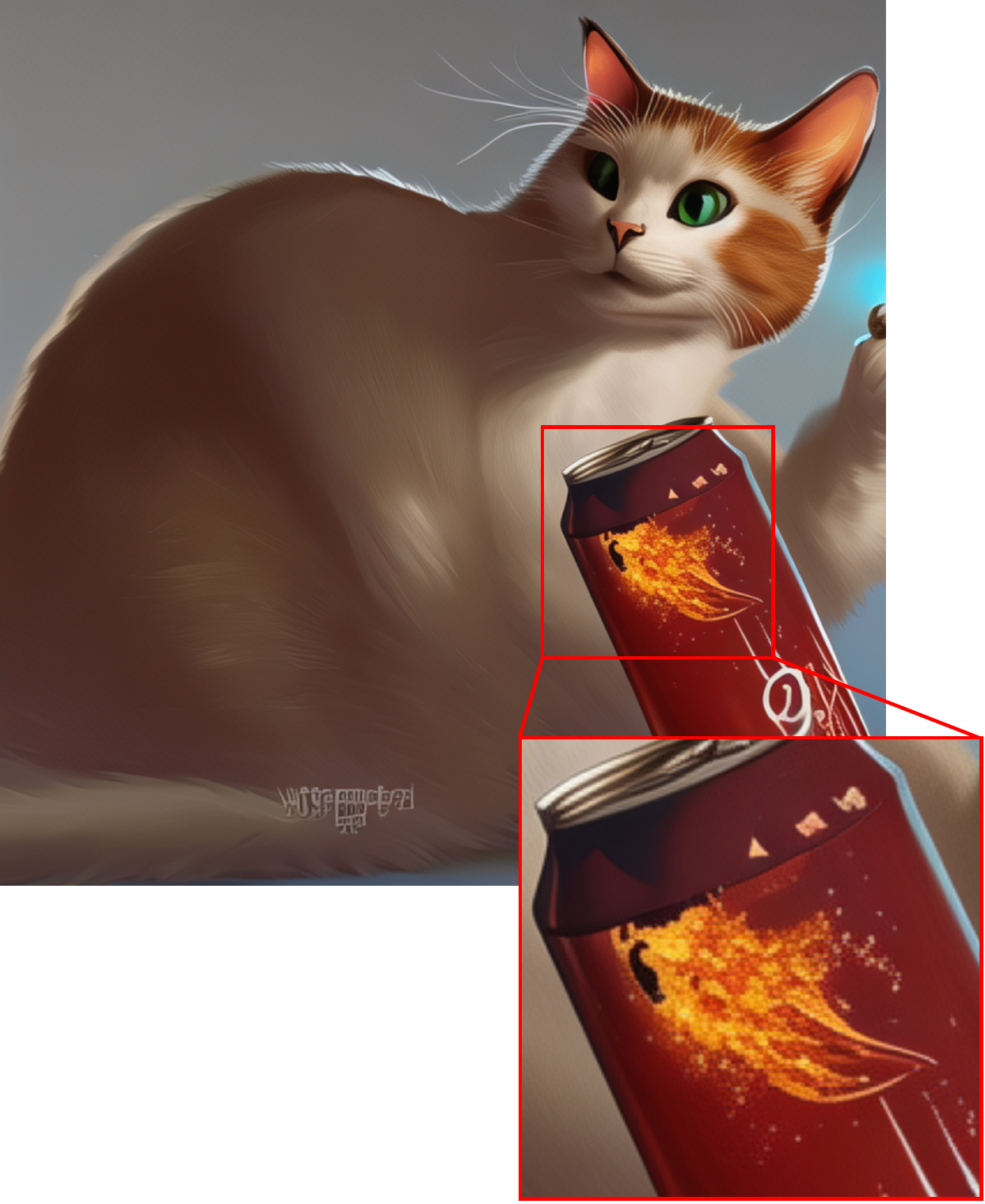}}
\end{subfigure}

\vspace{3mm}
\caption{An example of image generated by (a) the clean Stable Diffusion 2.1, (b) Stable Diffusion 2.1 watermarked by Stable Signature, (c) watermarked Stable Diffusion 2.1 fine-tuned by MP, (d) watermarked Stable Diffusion 2.1 fine-tuned by our attack with access to the encoder, and (e) watermarked Stable Diffusion 2.1 fine-tuned by our attack without access to the encoder. The same denoised latent vector is used by all diffusion models' decoders to generate the images. The watermark can only be detected in the image generated by (b). The image generated by (c) has significant loss of details. }
\label{fig-intro-comp}
  \vspace{-4mm}
\end{figure}

In this work, we propose a new model-targeted attack to remove in-generation watermark from open-source diffusion models. Our attack fine-tunes a diffusion model's decoder using a set of non-watermarked images, which we call  \emph{attacking dataset}. Specifically, our attack consists of two steps. In  Step I, we propose different methods to estimate a \emph{denoised latent vector}
for each non-watermarked image in the attacking dataset in two settings, i.e., with and without access to the diffusion model's encoder. The open-source diffusion model's decoder takes a  denoised latent vector  as input and outputs a watermarked image that is visually similar to the corresponding non-watermarked image. 
In Step II, we leverage the non-watermarked images in the attacking dataset and their corresponding estimated denoised latent vectors to fine-tune the diffusion model's decoder to remove the watermark from it. Our key idea is to fine-tune the decoder such that its generated images based on the denoised latent vectors are close to the corresponding non-watermarked images in the attacking dataset. 

We empirically evaluate our attack on the open-source diffusion model, i.e., Stable Diffusion 2.1, that is watermarked by Stable Signature. Our results show that our attack can effectively remove the watermark from the diffusion model such that its generated images are non-watermarked, while maintaining image quality. Moreover, our attack substantially outperforms MP,
the only existing model-targeted removal attack~\cite{fernandez2023stable}, in the scenario in which it is applicable. As shown in Figure~\ref{fig-intro-comp}, our attack can retain most information in the image after removing the watermark, while MP results in a blurry image with significant loss of details. Our results suggest that Stable Signature is not as robust as previously thought, and the design of a robust watermarking strategy for images generated by open-source diffusion models remains an open challenge.

\section{Related Works}
\subsection{Latent Diffusion Model}
\begin{figure}[t!]
\centering
{\includegraphics[width=\linewidth]{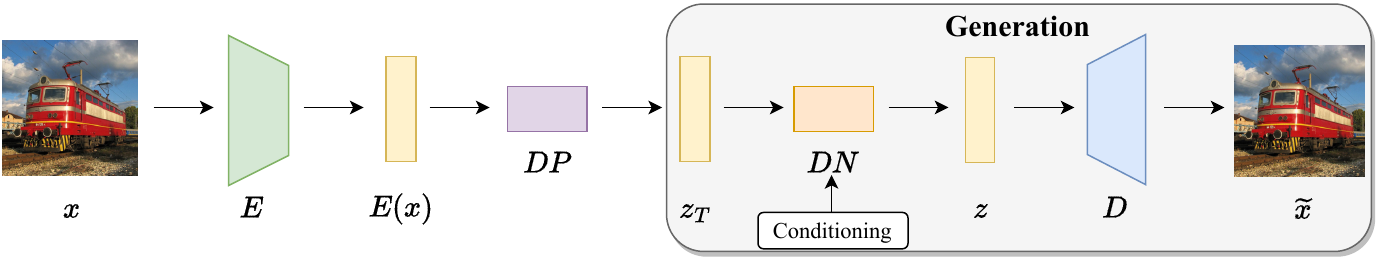}}
\vspace{3mm}
\caption{The main components of a latent diffusion model.}
\label{fig-ldm-overview}
\vspace{-4mm}
\end{figure}

Diffusion models~\cite{dhariwal2021diffusion, NEURIPS2020_4c5bcfec, NEURIPS2021_b578f2a5, ho2022cascaded} exhibit exceptional capability in generating images. A \emph{latent diffusion model}~\cite{rombach2022high} performs the diffusion process in the latent space, enhancing efficiency in both training of the diffusion model and image generation. 
A latent diffusion model has four main components: an encoder $E$ to encode an image $x$ into a \emph{latent vector} $E(x)$, diffusion process $DP$ to add Gaussian noise to the latent vector to obtain a \emph{noisy latent vector} $z_T=DP(E(x))$ where $T$ denotes the number of steps in diffusion process,  denoising layers $DN$ to obtain a \emph{denoised latent vector} $z=DN(z_T, c)$ where $c$ denotes the conditioning such as a text prompt or a depth map, and a decoder $D$ to reconstruct an image $D(z)$ from $z$. The diffusion process is a predefined probabilistic process that iteratively adds Gaussian noise to a latent vector, while the remaining three components are learnt using an image dataset. During image generation, a noisy latent vector $z_T'$ is sampled from Gaussian distribution, and the denoising layers $DN$ and decoder $D$ are used to generate an image $D(DN(z_T', c))$. The main components of a latent diffusion model are shown in Figure~\ref{fig-ldm-overview}.

\subsection{Image Watermark}
\myparatight{Post-generation watermark}
Post-generation watermarking methods~\cite{bi2007robust,al2007combined,zhu2018hidden,2019stegastamp,invisible-watermark,luo2020distortion, Jing_2021_ICCV} embed watermarks into images after the image generation process. These methods typically consist of three main components: a watermark (represented as a bitstring), a watermarking encoder for embedding the watermark into an image, and a watermarking decoder for extracting the watermark from an image. These methods can be categorized into two groups based on how the encoder and decoder are designed: \emph{learning-based} and \emph{non-learning-based}. Learning-based methods~\cite{zhu2018hidden,UDH,2019stegastamp,luo2020distortion} leverage deep learning techniques, utilizing neural networks for both encoding and decoding, while non-learning-based methods~\cite{pereira2000robust,al2007combined,bi2007robust,invisible-watermark} rely on manually crafted encoding and decoding algorithms. In closed-source setting, where the diffusion model is proprietary and users can only interact with it through API, learning-based watermarking methods exhibit significant robustness against various attacks~\cite{an2024benchmarking,2019stegastamp,jiang2023evading}. In open-source setting, however, such robustness is compromised. An attacker can easily remove the watermarking components from the open-source diffusion model, thus generating non-watermarked images without constraints.

\myparatight{Pre-generation watermark}
Pre-generation watermarking methods~\cite{wen2023tree} embed watermark into images before the image generation process. In diffusion models, for instance, a watermark can be incorporated into the noisy latent vector $z_T$~\cite{wen2023tree}. Subsequently, the image generated from this watermarked noisy latent vector contains the watermark. The watermark retrieval process involves an inverse operation of DDIM sampling~\cite{song2020improved}, which reconstructs the noisy latent vector from the generated image. However, such pre-generation watermark is also vulnerable in open-source setting. An attacker can substitute the watermarked noisy latent vector with a non-watermarked one, which is drawn from a Gaussian distribution. Consequently, image generated from this non-watermarked noisy latent vector does not contain the watermark.

\myparatight{In-generation watermark}
In-generation watermarking methods~\cite{fernandez2023stable} modify the parameters of the diffusion model's decoder to ensure that all images generated by this diffusion model inherently contain a watermark. This method seamlessly incorporates the watermarking components into the image generation process. For instance, Stable Signature~\cite{fernandez2023stable} fine-tunes the diffusion model's decoder using a HiDDeN~\cite{zhu2018hidden} watermarking decoder. After fine-tuning, each generated image carries a predetermined watermark which can be decoded by the watermarking decoder, i.e., directly incorporating the watermark into the diffusion model's parameters. This method is well-suited for open-source diffusion models, because it prevents attackers from removing the watermark by simply discarding the model's watermarking components.

\subsection{Watermark Removal Attacks}
\myparatight{Per-image-based}
Per-image-based removal attacks~\cite{jiang2023evading,an2024benchmarking,lukas2023leveraging,zhao2023invisible,saberi2023robustness} involve adding a carefully crafted perturbation on each watermarked image to remove the watermark. Common image processing techniques, such as JPEG compression and contrast adjustment, can introduce a perturbation for the watermarked image to remove the watermark. Furthermore, more sophisticated per-image-based removal attacks can be employed if the attacker has access to the watermarking decoder or detection API. For instance, Jiang et al.~\cite{jiang2023evading} proposed a white-box attack that assumes the attacker has access to the watermarking decoder, and a black-box attack that strategically manipulates the watermarked image based on detection API query results to remove the watermark.
These per-image-based removal attacks are applicable to all three groups of watermarks mentioned above as they do not require access to the image generation process.
However, they are inefficient when applied to a large volume of images due to the individualized design of perturbations for each watermarked image.

\myparatight{Model-targeted}
Model-targeted removal attacks~\cite{fernandez2023stable} are specifically designed for removing in-generation watermark. Such attacks involve modifying the diffusion model's parameters such that its generated images are non-watermarked. For instance, Stable Signature~\cite{fernandez2023stable} proposed MP to attack their Stable Signature in-generation watermark. This method aims to purify the diffusion model's decoder using non-watermarked images. However, it encounters challenges in effectively removing the watermark without significantly degrading image quality. Model-targeted removal attacks show high efficiency in removing watermark from numerous watermarked images, as it only requires a one-time modification of the diffusion model and images generated by the modified diffusion model are non-watermarked. These methods offer much higher efficiency compared to per-image-based removal attacks when handling numerous watermarked images.

\section{Problem Formulation}
\subsection{Watermarked Diffusion Model Decoder $D_w$}
We denote by $D_c$ a clean diffusion model decoder without watermark. $D_c$ is fine-tuned as a watermarked diffusion model decoder $D_w$  such that its generated images are inherently embedded with a ground-truth watermark $w_g$. Formally, any generated image $D_w(DN(z_T, c))$ is embedded with $w_g$, where $z_T$ is a noisy latent vector sampled from a Gaussian distribution, $DN$ is the denoising layers, and $c$ is the conditioning. $D_w$ is made open-source, allowing users to generate watermarked images. 

\subsection{Threat Model}
\myparatight{Attacker's goals} Given a watermarked diffusion model decoder $D_w$, an attacker aims to fine-tune it as a non-watermarked diffusion model decoder $D_{nw}$. Specifically, the attacker aims to achieve two goals: 1) \emph{effectiveness goal}, and 2) \emph{utility goal}. The effectiveness goal means that images generated by $D_{nw}$ do not have the watermark $w_g$ embedded; while the utility goal means that the images generated by  $D_{nw}$ maintain visual quality, compared to those generated by $D_w$. 

\myparatight{Attacker's knowledge} 
A watermarked latent diffusion model consists of an encoder $E$, diffusion process $DP$,  denoising layers $DN$, and a watermarked decoder $D_w$.
The denoising layers $DN$ and decoder $D_w$ are involved when generating images, i.e., $D_w(DN(z_T, c))$ is a generated image, where $z_T$ is a noisy latent vector sampled from Gaussian distribution and $c$ is the conditioning. We assume $DN$ and $D_w$ are open-source, and thus the attacker has access to them. Depending on whether $E$ and $DP$ are open-source, we consider the following two scenarios:
\begin{itemize}
    \item {\bf Encoder-aware (E-aware).}
    In this scenario, the model provider also makes $E$ and $DP$ open-source. Therefore, the attacker has access to them. For instance, Stable Diffusion model makes its $E$ and $DP$ open-source.
    \item {\bf Encoder-agnostic (E-agnostic).}
    In this scenario, $E$ and $DP$ are not open-source, e.g., because image generation only requires $DN$ and $D_w$. Therefore, the attacker does not have access to $E$ and $DP$ in this setting.
\end{itemize}

Additionally, we assume the attacker has access to a set of non-watermarked images, which we call attacking dataset. For instance, the attacker can simply use popular benchmark images (e.g., ImageNet) as the attacking dataset. The attacking dataset is used to remove watermark from the watermarked diffusion model decoder $D_w$.  

\myparatight{Attacker's capability}
We assume the attacker can  modify the parameters of the open-sourced watermarked latent diffusion model decoder $D_w$. The denoising layers $DN$, which are much larger than the decoder, requires much more computational resources to modify. For instance, in Stable Diffusion 2.1, the denoising layers have about 10 times more parameters than the decoder. Therefore, we assume the attacker modifies the decoder. 
\section{Our Attack}
\subsection{Overview}
\begin{figure}[t!]
\centering
{\includegraphics[width=0.7\linewidth]{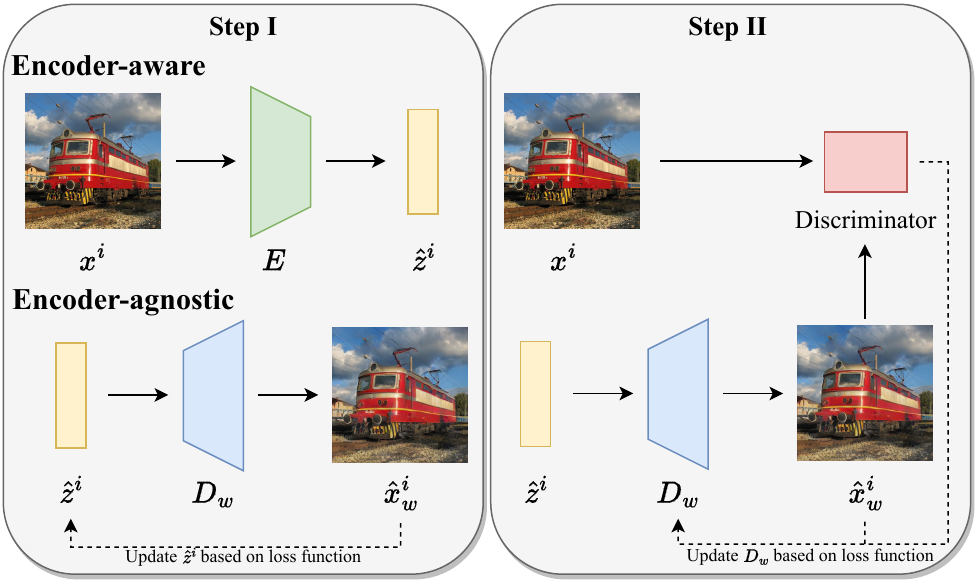}}
\vspace{3mm}
\caption{Overview of our attack. The solid arrows represent the direction of data flow and the dashed arrows represent the direction of gradient flow.}
\label{fig-attk-overview}
\end{figure}

We propose a two-step method to fine-tune the decoder $D_w$ to make the diffusion model's generated images non-watermarked using an attacking dataset of size $n$, as illustrated in Figure~\ref{fig-attk-overview}. In Step I, we estimate the denoised latent vector $z^i$ for each non-watermarked image $x^i$ in the attacking dataset, where $i=1,2,\ldots,n$. In Step II, by utilizing these images and their estimated denoised latent vectors, we fine-tune the decoder $D_w$ to ensure that the reconstructed images closely match the non-watermarked images when the inputs are the corresponding estimated denoised latent vectors. 
Our intuition is that a watermarked decoder will transform a denoised latent vector $z^i$ to the watermarked version of $x^i$, denoted as $x^i_w$. Therefore, through fine-tuning the decoder to reconstruct $x^i$ from the input $z^i$, the decoder is trained to map any given denoised latent vector to the non-watermarked version of its corresponding image, effectively removing watermarks from images generated by the diffusion model.

\subsection{Step I: Estimate the Denoised Latent Vector $z$}
To estimate the denoised latent vector $z^i$ for the non-watermarked image $x^i$, we propose different methods in different scenarios.

\myparatight{E-aware} 
In this scenario, an attacker has access to the encoder $E$, diffusion process $DP$, denoising layers $DN$, and watermarked decoder $D_w$.
Based on the pipeline of the diffusion model, the denoised latent vector $z^i$ can be represented as $z^i=DN(DP(E(x^i)), c^i)$. However, since we don't have access to the ground-truth conditioning $c^i$ to reconstruct $z^i$, we cannot directly compute $z^i$ even though we have access to $E$, $DP$, and $DN$.
We observe that the denoising layers $DN$ are trained to denoise the noisy latent vector $z_T$ such that $DN(z_T, c)$ is close to $E(x)$. Therefore, the attacker can utilize the encoder to encode the non-watermarked image $x^i$ into the latent space to get an estimation of the denoised latent vector $z^i$, denoted by $\hat{z}^i$, as follows:
\begin{equation}
\begin{aligned}
\hat{z}^i = E(x^i), \forall i.
\end{aligned}
\label{eqn-0}
\end{equation}

\myparatight{E-agnostic}
In this scenario, an attacker only has access to the denoising layers $DN$ and watermarked decoder $D_w$. The most straightforward way to estimate the denoised latent vector $z^i$ is to train a new encoder based on $DN$ and $D_w$ and use the method in E-aware scenario. However, training an encoder from scratch for a latent diffusion model to achieve good encoding performance requires a large number of data and computational resources, which is very time-consuming and infeasible for an attacker with limited resources. Recall that our goal is to estimate the denoised latent vector $z^i$ which will be mapped to the watermarked image $x^i_w$ by the watermarked decoder $D_w$. Formally, we can formulate an equation as follows:

\begin{equation}
\begin{aligned}
D_w(z^i) = x^i_w, \forall i.
\end{aligned}
\label{eqn-1}
\end{equation}

This equation is difficult to solve since there are two variables in it, the denoised latent vector $z^i$ and watermarked image $x^i_w$. To reduce the number of variables, we use the known $x^i$ as an approximation of $x^i_w$ since the watermarked version of an image should be highly perceptually close to the non-watermarked version. Therefore, to get an estimation of $z^i$, we can reformulate the equation as follows:

\begin{equation}
\begin{aligned}
D_w(\hat{z}^i) = x^i, \forall i.
\end{aligned}
\label{eqn-2}
\end{equation}

We can easily get an estimation of $z^i$ for Equation~\ref{eqn-2} if $D_w$ is invertible, i.e., $\hat{z}^i = D_w^{-1}(x^i), \forall i$. However, since the diffusion model's decoder is a complicated neural network and it is usually infeasible to get its inverse function, solving the Equation~\ref{eqn-2} directly is challenging. To address the challenge, we can treat $\hat{z}^i$ as a trainable variable and reformulate Equation~\ref{eqn-2} into an optimization problem as follows:
\begin{equation}
\begin{aligned}
& \min _{\hat{z}^i} l_p(D_w(\hat{z}^i), x^i), \forall i,
\end{aligned}
\label{optim-1}
\end{equation}
where $l_p(\cdot,\cdot)$ denotes the perceptual loss between two images to ensure the visual similarity. However, it is still challenging to make $D_w(\hat{z}^i)$ closely resemble the non-watermarked image $x^i$ since $\hat{z}^i$ is randomly initialized and $D_w(\hat{z}^i)$ is completely different from $x^i$ at the early stage of the optimization process.

Therefore, we propose a two-stage optimization method to solve the optimization problem described in Equation~\ref{optim-1}. At the first stage, for each $\hat{z}^i$, we randomly initialize it using a standard Gaussian distribution. Then we employ gradient descent to find an initial point $\hat{z}^i_{init}$ for $\hat{z}^i$ that minimizes the mean square error between $D_w(\hat{z}^i_{init})$ and $x^i$. This stage ensures that $D_w(\hat{z}^i_{init})$ roughly resembles $x^i$, though with a significant loss of detailed information. At the second stage, we initialize $\hat{z}^i$ with the initial point $\hat{z}^i_{init}$ obtained from the first stage. Then we set $l_p(\cdot,\cdot)$ to be the Watson-VGG perceptual loss introduced by Czolbe et al.~\cite{czolbe2020loss} and use gradient descent to further optimize $\hat{z}^i$, enabling it to capture and reconstruct the detailed information of the non-watermarked image $x^i$. The detailed method to estimate the denoised latent vector $z^i$ in E-agnostic scenario is shown in Algorithm~\ref{alg_1} in Appendix.

\subsection{Step II: Fine-tune the Decoder $D_w$}
Given a set of estimated denoised latent vectors $\hat{z}^i$ and non-watermarked images $x^i$, our goal is to modify the parameters of the watermarked decoder $D_w$ to make the diffusion model's generated images non-watermarked. The main idea is to modify the decoder's parameters to enable it to map the denoised latent vector $z^i$, which is originally mapped to the watermarked image $x^i_w$, to the non-watermarked image $x^i$. To achieve this, we use the estimated denoised latent vectors $\hat{z}^i$ and non-watermarked images $x^i$ to fine-tune the decoder, ensuring that the reconstructed images closely resemble the non-watermarked images at the pixel level to effectively remove the watermark signal from each pixel. Formally, we can formulate the optimization problem as follows:
\begin{equation}
\begin{aligned}
& \min _{D_w} \frac{1}{n} \sum_{i=1}^{n}\| D_w(\hat{z}^i) - x^i \|_{2}.
\end{aligned}
\label{optim-2}
\end{equation}

However, since the mean square error measures the average difference between the non-watermarked and reconstructed images, it tends to penalize large errors more severely than small ones, leading to a smoothing effect where the reconstructed images may lose lots of detailed information. To solve this challenge, a perceptual loss that measures the distance of the high-level features produced by a pre-trained neural network between two images is employed to ensure the visual quality of the reconstructed images. Formally, we can reformulate the optimization problem as follows:
\begin{equation}
\begin{aligned}
& \min _{D_w} \frac{1}{n} \sum_{i=1}^{n} \| D_w(\hat{z}^i) - x^i \|_{2} + \lambda \frac{1}{n} \sum_{i=1}^{n} l_p(D_w(\hat{z}^i), x^i),
\end{aligned}
\label{optim-3}
\end{equation}
where $\lambda$ denotes the weight for the perceptual loss. To solve the optimization problem, we employ gradient descent to optimize the parameters of $D_w$ to minimize the objective function in Equation~\ref{optim-3}. During the optimization, we adopt a convolution neural network introduced by Zhu et al.~\cite{zhu2018hidden} as a discriminator to perform adversarial training. The discriminator is trained to distinguish $D_w(\hat{z}^i)$ from $x^i$ and the decoder $D_w$ is trained to fool the discriminator. Formally, we reformulate the optimization problem as follows:
\begin{equation}
\begin{aligned}
\min _{D_w} & \frac{1}{n} \sum_{i=1}^{n} \| D_w(\hat{z}^i) - x^i \|_{2} + \lambda \frac{1}{n} \sum_{i=1}^{n} l_p(D_w(\hat{z}^i), x^i) \\
& \quad + \mu \frac{1}{n} \sum_{i=1}^{n} log(1-disc(D_w(\hat{z}^i))),
\end{aligned}
\label{optim-4}
\end{equation}
where $disc$ denotes the discriminator and $\mu$ denotes the weight for the adversarial loss. The detailed method to fine-tune the decoder $D_w$ is shown in Algorithm~\ref{alg_2} in Appendix.
\section{Evaluation}
\subsection{Experimental Setup}
\myparatight{Datasets}
We employ public non-AI-generated images as our attacking datasets. Specifically, we utilize three datasets: ImageNet~\cite{ILSVRC15}, MS-COCO~\cite{lin2014microsoft}, and Conceptual Captions~\cite{sharma-etal-2018-conceptual}. From each dataset, we randomly select 4,000 images as an attacking dataset to fine-tune the watermarked decoder. The images in the attacking datasets are resized to 256 $\times$ 256. For testing, we evaluate the effectiveness and utility goals using images generated by an open-source watermarked diffusion model and its versions fine-tuned by watermark removal attacks. These images are produced using text prompts from the Stable Diffusion Prompts dataset created by MagicPrompt~\cite{magic-prompt}. Specifically, we randomly sample 100 text prompts from the dataset to generate 100 images for testing.

\myparatight{Detecting watermark in an image} A watermarking decoder $W_d$ is used to detect whether $w_g$ is in an image $x$. Specifically,   $W_d$ is used to decode a watermark,  represented as $W_d(x)$, from the image $x$. The bitwise accuracy $BA(w_1,w_2)$ between two watermarks $w_1$ and $w_2$ is the proportion of bits that are identical in $w_1$ and $w_2$. $x$ is detected as  watermarked with $w_g$ if the bitwise accuracy $BA(W_d(x), w_g)$ exceeds a detection threshold $\tau$ or falls below $1-\tau$, i.e., $BA(W_d(x), w_g)>\tau$ or $BA(W_d(x), w_g)<1-\tau$. Such detector is known as \emph{double-tail detector}~\cite{jiang2023evading}, which is more robust than \emph{single-tail detector} that detects the image $x$ as watermarked if the bitwise accuracy $BA(W_d(x), w_g)$ exceeds $\tau$. Therefore, we use double-tail detector in this work. 

\myparatight{Diffusion model and watermarking decoder}
We use the open-source clean Stable Diffusion 2.1 and its watermarked version obtained by Stable Signature~\cite{fernandez2023stable}. For the watermarked version, the watermarked decoder $D_w$ is fine-tuned from the Stable Diffusion 2.1's clean decoder $D_c$ with the MS-COCO dataset by Stable Signature. The images generated by the watermarked Stable Diffusion 2.1 are embedded with a ground-truth watermark $w_g$ that has 48 bits. For the watermarking decoder $W_d$, we use the open-source one~\cite{fernandez2023stable} in Stable Signature, which aims to detect whether $w_g$ is embedded in an image.

\myparatight{Different variants to estimate the denoised latent vector $z$}
In our experiments, we compare our two-stage optimization method (denoted by 2S) with the following variants to estimate the denoised latent vector $z$. All of these methods initialize $\hat{z}$ with a standard Gaussian distribution and treat it as a trainable variable.
\begin{itemize}
    \item {\bf One-stage mean square error (1S-M)}
    This method optimizes $\hat{z}$ to minimize the mean square error between the reconstructed image $D_w(\hat{z})$ and the non-watermarked image $x$.

    \item {\bf One-stage perceptual loss  (1S-P)}
    This method optimizes $\hat{z}$ to minimize the perceptual loss calculated by the Watson-VGG model between $D_w(\hat{z})$ and $x$.

    \item {\bf One-stage mixed loss  (1S-Mix)}
    This method optimizes $\hat{z}$ to minimize the mixed loss consisting of mean square error and perceptual loss calculated by the Watson-VGG model between $D_w(\hat{z})$ and $x$. The weights for different loss functions are set to be 1.
    
\end{itemize}

\myparatight{Per-image-based removal attacks}
In our experiments, we compare our attack with five commonly used per-image-based removal attacks, including the state-of-the-art one proposed by Jiang et al.~\cite{jiang2023evading}. The details of the per-image-based removal attacks we use are shown in Appendix~\ref{apdx-per-image}.
It should be emphasized that all of these per-image-based attacks require to craft a perturbation for each watermarked image individually to remove watermark.

\myparatight{Model-targeted removal attack}
For model-targeted attacks, we compare our attack with MP introduced in Stable Signature~\cite{fernandez2023stable}. Specifically, MP involves fine-tuning the diffusion model's encoder and decoder with the encoder's parameters fixed to reconstruct non-watermarked images using mean square error as the reconstruction loss. Note that this method requires the access to the diffusion model's encoder and is only applicable in the E-aware scenario. The parameter setting follows the configuration by Stable Signature~\cite{fernandez2023stable} as shown in Appendix~\ref{apdx-mp}.

\myparatight{Evaluation metrics}
To evaluate whether our attack achieves the effectiveness goal, we utilize two metrics: \emph{evasion rate} and \emph{bitwise accuracy}. Evasion rate is the proportion of generated images (or perturbed images for per-image-based removal attacks) detected as watermarked by the  watermark-based detector. Bitwise accuracy is the proportion of bits in the watermark decoded from a generated (or perturbed) image that matches with $w_g$. Additionally, to evaluate whether our attack achieves the utility goal, we use a commonly used metric for the generation quality of generative models, i.e., \emph{Fréchet Inception Distance (FID)}. Specifically, we compute the FID in the testing set between generated (or perturbed) images and the watermarked images generated by the watermarked Stable Diffusion 2.1 with the same random seed. Note that the bitwise accuracy is averaged across 100 images in the testing set.

\myparatight{Parameter settings}
In the E-aware scenario, we use the Watson-VGG~\cite{czolbe2020loss} model to measure the perceptual loss in Step II.
 However, in Step I of our attack, we use the Watson-VGG model to measure the perceptual loss in the E-agnostic scenario. 
To avoid potential local minima issues that could emerge from using the same perceptual loss model, we  use VGG-16~\cite{torch-vgg} to measure the perceptual loss in E-agnostic scenario in Step II. For the discriminator $disc$, we employ the discriminator in HiDDeN~\cite{zhu2018hidden}.

To estimate the denoised latent vector $z$ in the E-agnostic scenario, 2S is employed as the default method and we execute 500 epochs for each stage. In each stage, the Adam optimizer, with a learning rate of 0.1, is used to optimize $\hat{z}$. For other variants to estimate $z$, we execute 1,000 epochs--equivalent to the total epoch count in 2S--and maintain consistent optimizer settings.

For decoder fine-tuning, we execute 1 epoch in the E-aware scenario and 2 epochs in the E-agnostic scenario. We set the parameters $\lambda=1$ and $\mu=0.1$. Additionally, the AdamW optimizer is used, with a base learning rate of 0.0005 with a linear warm-up period of 20 iterations followed by a half-cycle cosine decay. The batch size is set to be 4. For optimizing the discriminator, the Adam optimizer is used with a learning rate of 0.001.

The detection threshold $\tau$ is set to ensure that the false positive rate of the double-tail detector does not exceed $10^{-4}$. Given that the watermark length in our experiments is 48, $\tau$ is set to be 0.77.

\subsection{Experimental Results}
\vspace{-2mm}
\myparatight{Our attack achieves both the effectiveness and utility goals}
\begin{figure}[t!]
    \centering
    \subfloat[Evasion rate]{\includegraphics[width=0.33 \textwidth]{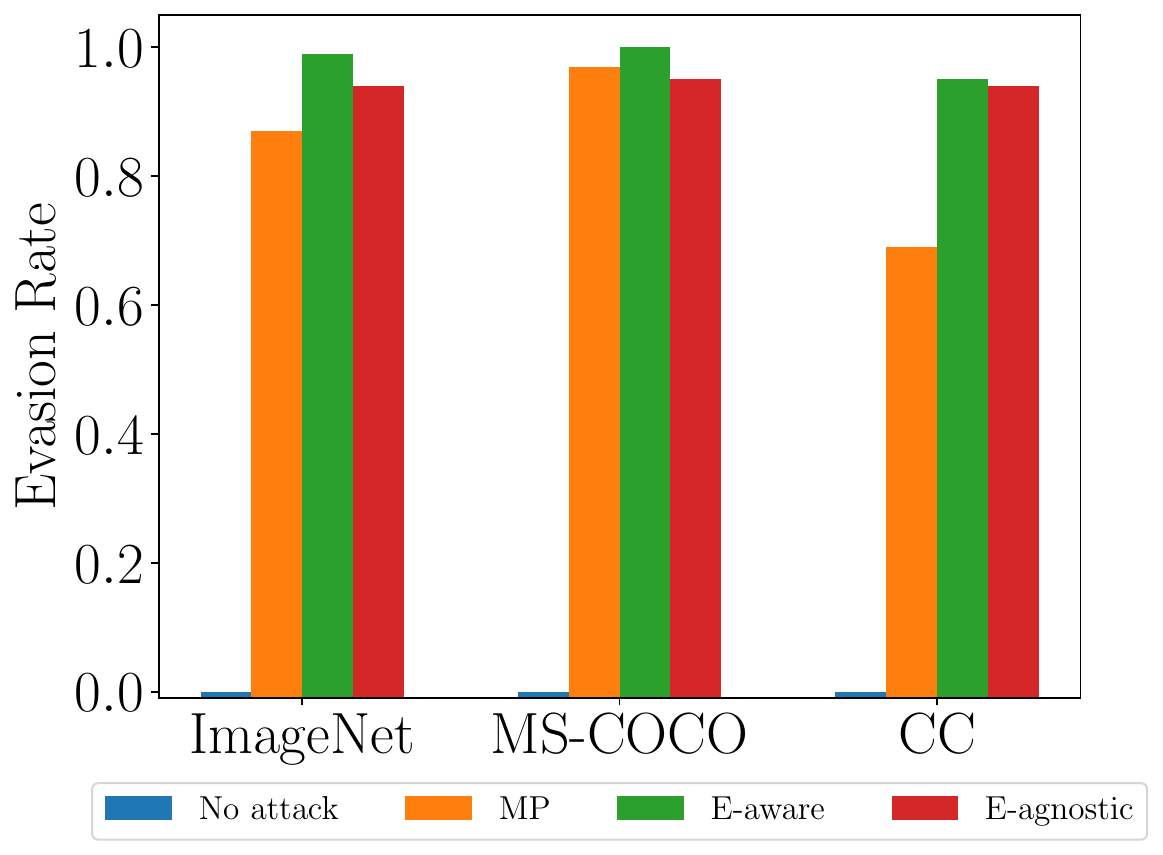}}
    \subfloat[Bitwise accuracy]{\includegraphics[width=0.33 \textwidth]{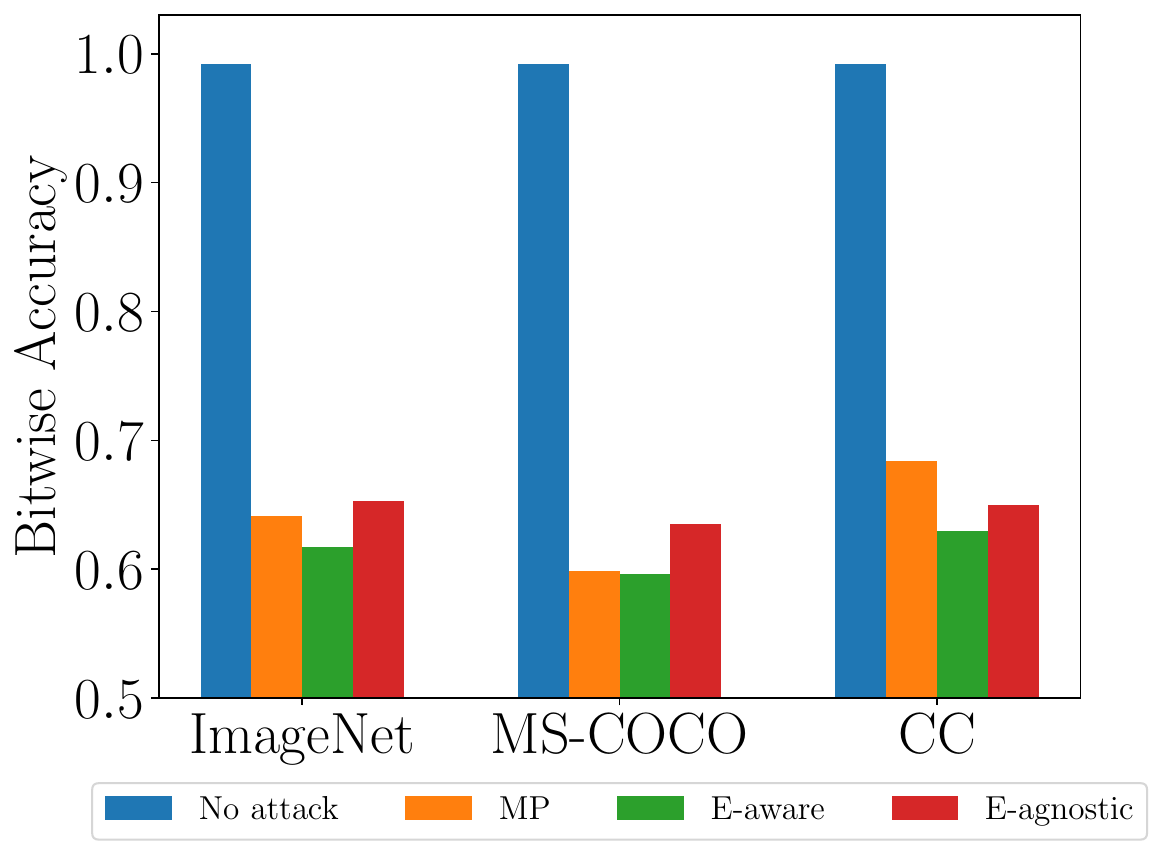}}
    \subfloat[FID]{\includegraphics[width=0.33 \textwidth]{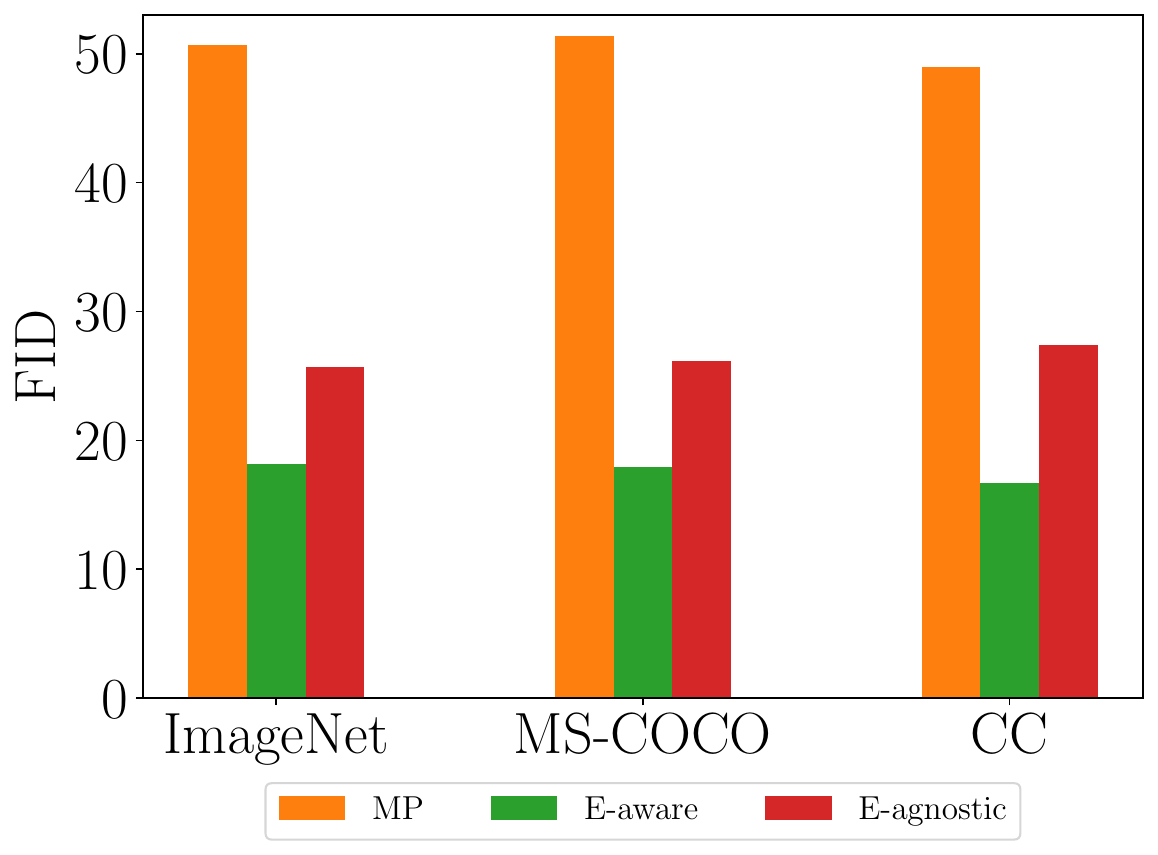}}
    \vspace{3mm}
    \caption{Effectiveness and utility of  MP and our attack with the three attacking datasets.}
\label{fig-main}
\end{figure}
Figure~\ref{fig-main} shows the evasion rate, bitwise accuracy, and FID of  MP and our attack with the three attacking datasets. First, we observe that our attack is successful in evading watermark-based detection in both E-aware and E-agnostic scenarios. The evasion rate is higher than 94\% and the bitwise accuracy is lower than 66\%, while maintaining an FID lower than 27.5. Second, we observe that our attack outperforms MP in both scenarios. In the E-aware scenario, our attack has a higher evasion rate and lower bitwise accuracy while maintaining a much lower FID in all the three attacking datasets. In the E-agnostic scenario, our attack still has a comparable or higher evasion rate and comparable bitwise accuracy while maintaining a much lower FID in all the three attacking datasets. Note that MP assumes the attacker has access to the encoder while our attack in the E-agnostic scenario doesn't. Figure~\ref{fig-comp} in Appendix shows some image examples of our attack compared with the clean and watermarked ones. We observe that the images produced by our non-watermarked decoder are almost the same as those produced by the clean and watermarked decoder.

\myparatight{Comparing with per-image-based removal attacks}
\begin{table}[t!]
    \centering
        \caption{Utility and processing time of per-image-based attacks and our attack.}
        \vspace{3mm}
    \begin{tabular}{|c|c|c|c|c|c|} \hline
        {} & \multicolumn{3}{c|}{Utility} & \multicolumn{2}{c|}{Time} \\ \hline
        {Method} & {FID $\downarrow$} & {PSNR $\uparrow$} & {SSIM $\uparrow$} & {Fine-tuning (min) $\downarrow$} & {Removal (s/img) $\downarrow$} \\ \hline
        {JPEG} & {61.48} & {28.57} & {0.84} & {0} & {0.036}  \\ \hline
        {Brightness} & {223.67} & {5.15} & {0.40} & {0} & {0.005}  \\ \hline
        {Contrast} & {172.44} & {10.10} & {0.32} & {0} & {0.002}  \\ \hline
        {GN} & {228.51} & {13.10} & {0.11} & {0} & {0.017}  \\ \hline
        {WEvade-W-II} & {5.80} & {38.94} & {0.99} & {0} & {651.034}  \\ \hline
        {E-aware} & {18.15} & {29.50} & {0.86} & {14.197} & {0}  \\ \hline
        {E-agnostic} & {25.68} & {29.40} & {0.86} & {8777.885} & {0}  \\ \hline
    \end{tabular}
    \label{tab-per-image}
\end{table}
Table~\ref{tab-per-image} shows the utility and the processing time of our attack compared with five per-image-based removal attacks when achieving  similar evasion rate and bitwise accuracy. Figure~\ref{fig-per-image} in Appendix shows the comparison of the generated (or perturbed) images by different attacks. We also show the \emph{Peak signal-to-noise ratio (PSNR)} and \emph{Structural Similarity Index Measure (SSIM)} which are the commonly used metrics for assessing per-image-based attacks' utility. For the processing time, we divide it into two phases: decoder fine-tuning and watermark removal for each image. It is important to note that the fine-tuning time of our attack is on a single NVIDIA A6000 GPU. It can be significantly reduced when using multiple GPUs since the process of estimating $z$ which is the most time-consuming part in E-agnostic scenario can be parallelized. For instance, with four NVIDIA A6000 GPUs, it only takes about 2K minutes for fine-tuning in E-agnostic scenario. 

First, we observe that the utility of our attack is much higher than most per-image-based removal attacks. Second, the removal time of our attack is 0 once the decoder is fine-tuned. Therefore, our attack shows much higher efficiency of removing watermark when the number of generated images is large. For instance, the efficiency of our attack outperforms WEvade-W-II when processing more than 1 image in E-aware scenario and 809 images in E-agnostic scenario. Note that WEvade-W-II requires the access to the watermarking decoder $W_d$ to perform a white-box attack, and it represents the upper bound of the utility that can be achieved by a removal attack. Though our attack maintains a slightly worse utility than WEvade-W-II when compared to the watermarked images, it is still difficult for human's eyes to notice their differences with the watermarked images, as shown in Figure~\ref{fig-per-image} in Appendix. It is worthwhile to mention that our attack maintains a similar utility as WEvade-W-II when compared to the clean non-watermarked images. It is because we optimize the decoder's output close to the non-watermarked image rather than the watermarked one in the optimization problem we formulate.

\myparatight{Different variants to estimate $z$}
\begin{figure}[t!]
\centering
      \raisebox{-0.5\height}{\subfloat[FID]{\includegraphics[width=.18\linewidth]{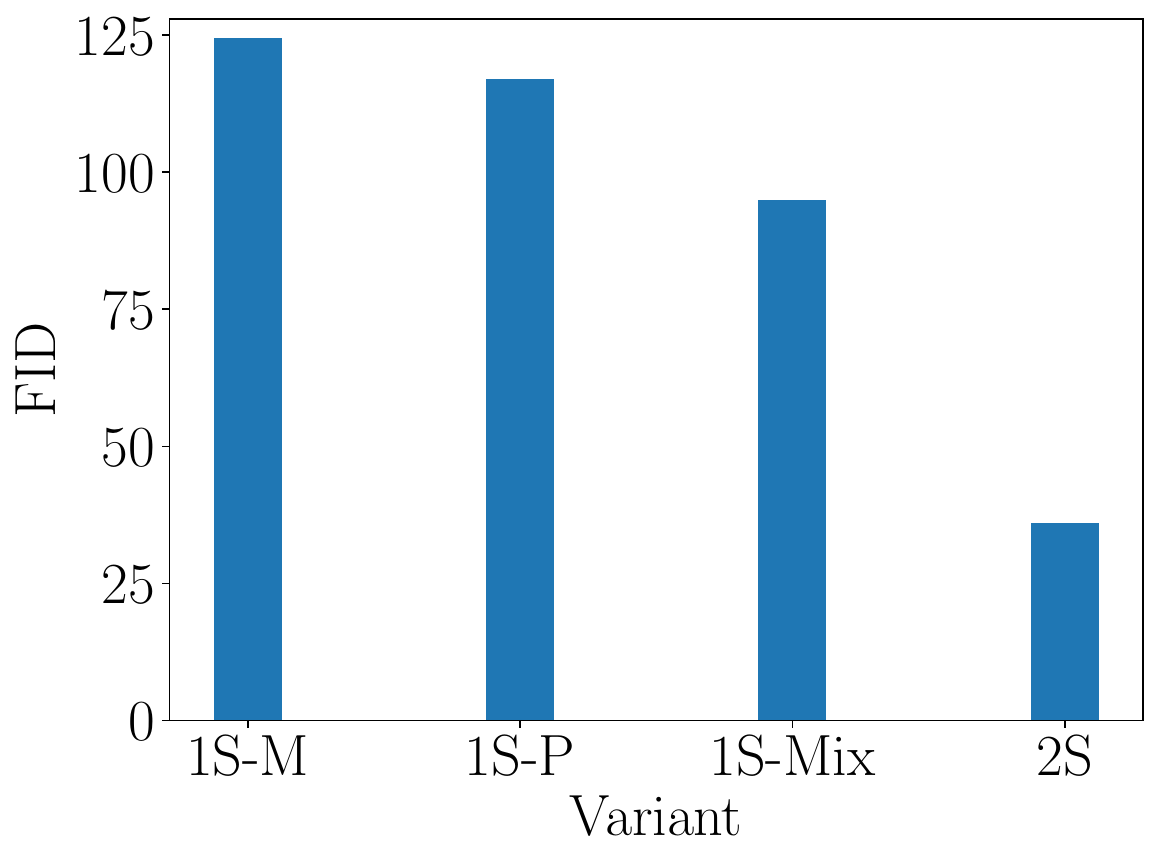}}}
      \raisebox{-0.5\height}{\subfloat[NW]{\includegraphics[width=.13\linewidth]{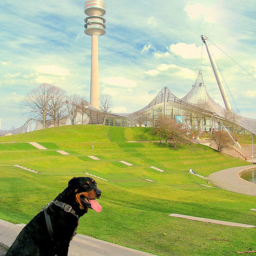}}}
      \raisebox{-0.5\height}{\subfloat[1S-M]{\includegraphics[width=.13\linewidth]{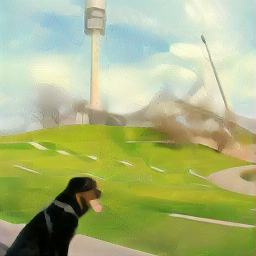}}}
      \raisebox{-0.5\height}{\subfloat[1S-P]{\includegraphics[width=.13\linewidth]{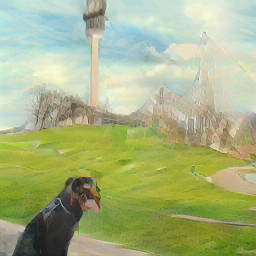}}}
      \raisebox{-0.5\height}{\subfloat[1S-Mix]{\includegraphics[width=.13\linewidth]{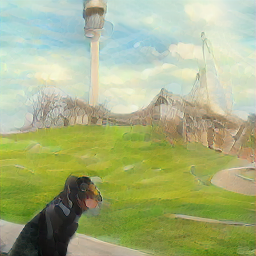}}}
      \raisebox{-0.5\height}{\subfloat[2S]{\includegraphics[width=.13\linewidth]{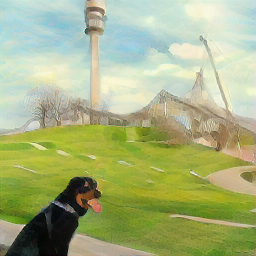}}}
\vspace{3mm}
\caption{Image reconstruction performance for different variants to estimate $z$ on ImageNet. NW denotes the non-watermarked image.}
\label{fig-zi}
\end{figure}
Figure~\ref{fig-zi} shows the FID and the examples of reconstructed image by different variants to estimate the denoised latent vector $z$ for a non-watermarked image. The FID is computed between 100 images randomly selected from ImageNet dataset and their reconstructed version by different variants. We observe that the images reconstructed by 2S are more similar to the original ones than those reconstructed by other variants. 2S achieves a much lower FID compared with other variants. Additionally, from the examples of reconstructed image shown, we observe that $z$ produced by our method can reconstruct more detail information in the original image and achieve a higher level of visual similarity to the original one.

\myparatight{Different $\lambda$}
\begin{figure}[t!]
    \centering
    \subfloat[Evasion rate]{\includegraphics[width=0.33 \textwidth]{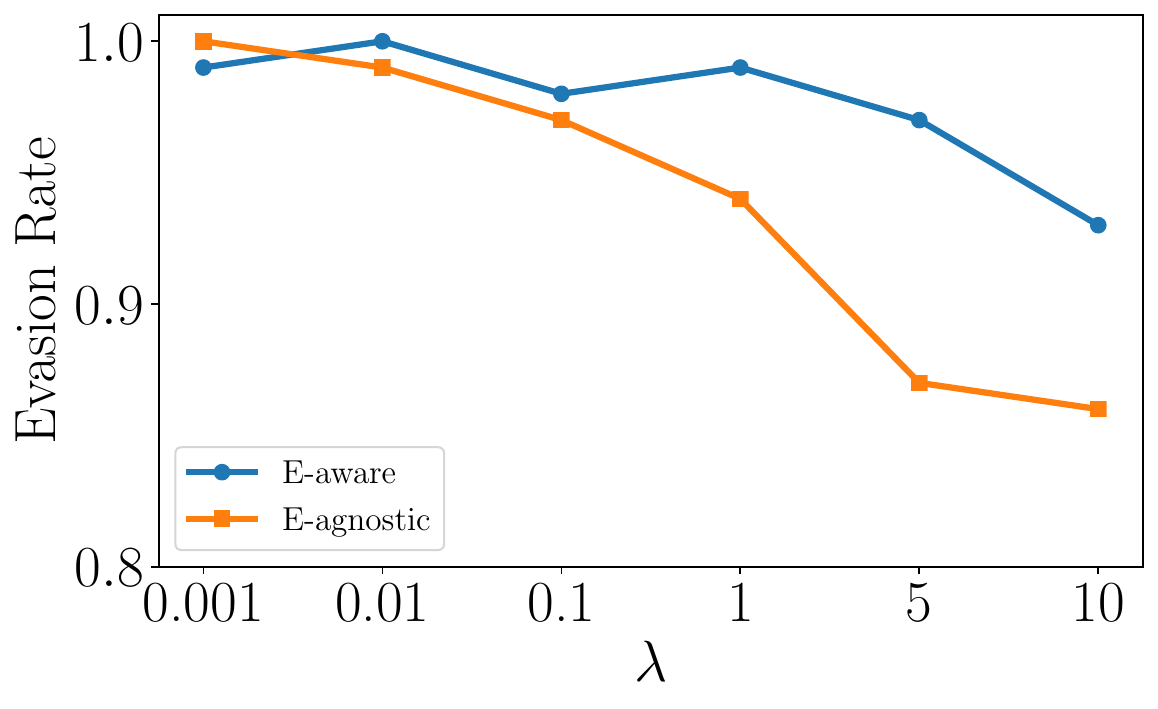}}
    \subfloat[Bitwise accuracy]{\includegraphics[width=0.33 \textwidth]{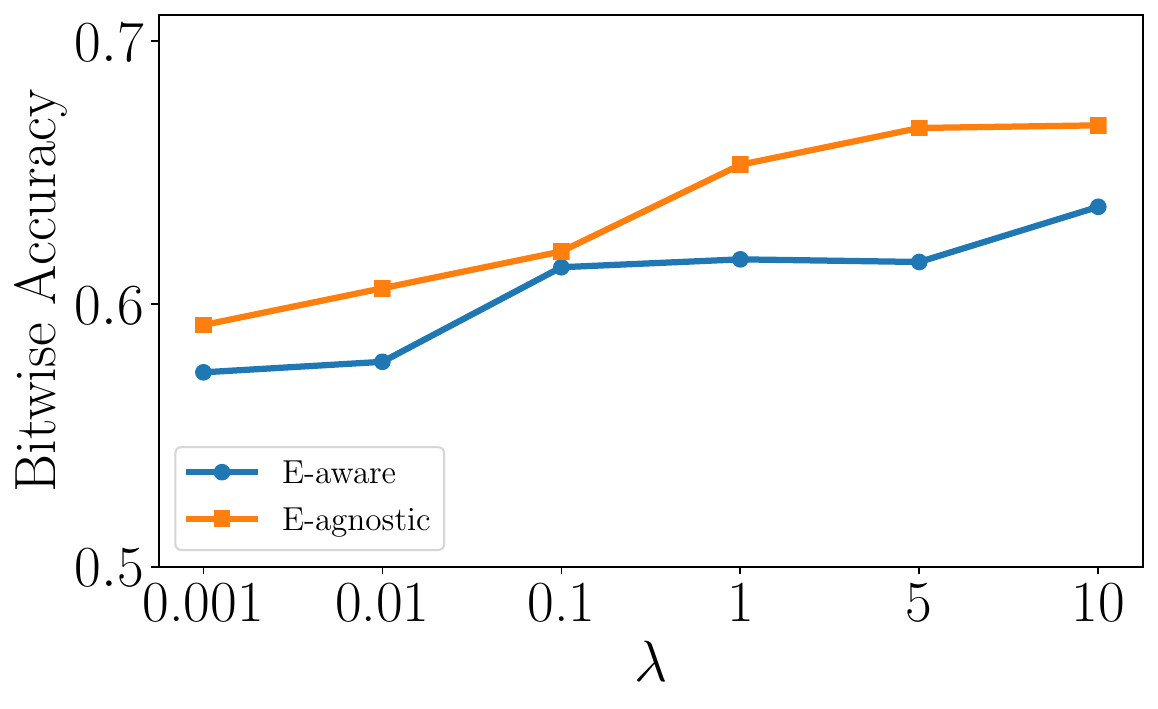}}
    \subfloat[FID]{\includegraphics[width=0.33 \textwidth]{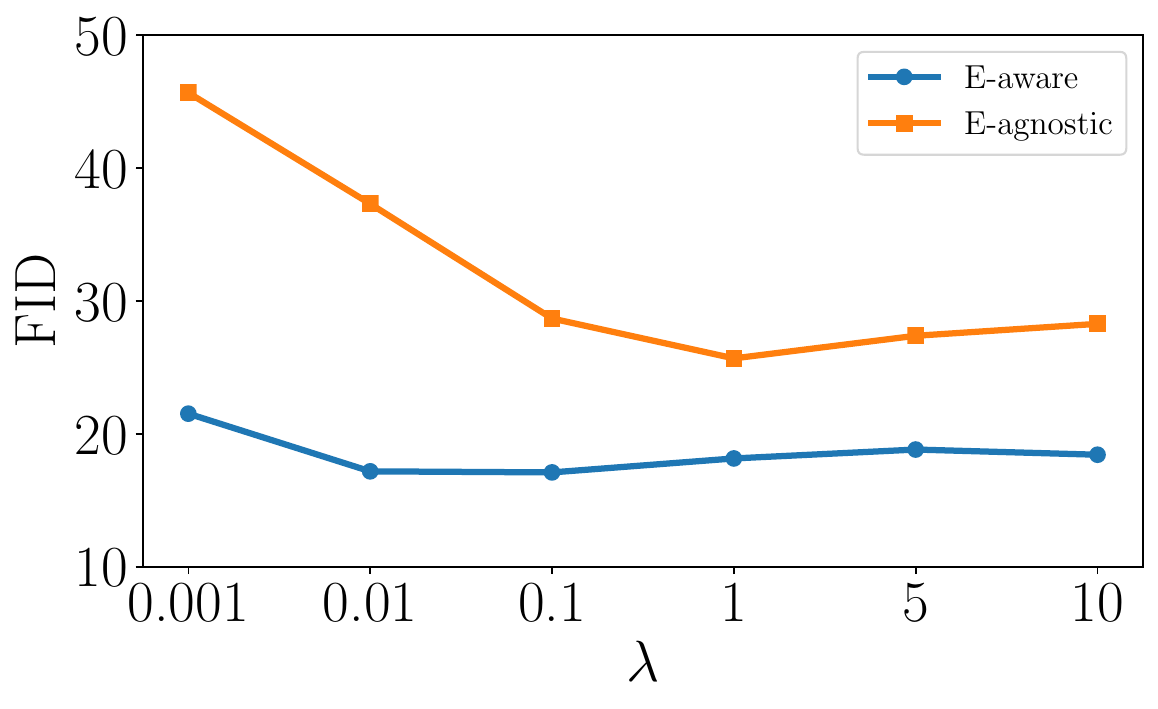}}
    \vspace{3mm}
    \caption{Effectiveness and utility of our attack with different $\lambda$ on ImageNet.}
\label{fig-lambda}
\end{figure}
Figure~\ref{fig-lambda} shows the evasion rate, bitwise accuracy, and FID for different $\lambda$ in our attack. First, we observe that the effectiveness of our attack decreases when $\lambda$ increases. This trend can be attributed to the loss function placing greater emphasis on perceptual loss, thereby reducing the mean square error's capability in removing watermarks. Second, we observe that the utility of our attack first increases and then slightly decreases when $\lambda$ increases. The early improvement in utility results from a higher weighting on perceptual loss, which enhances image fidelity. However, when $\lambda$ continues increasing, the reconstructed image starts to deviate from the non-watermarked image pixel-wisely since the loss function focus more on the perceptual loss than the mean square error, which results in a worse utility.

\myparatight{Different $\mu$}
\begin{figure}[t!]
    \centering
    \subfloat[Evasion rate]{\includegraphics[width=0.33 \textwidth]{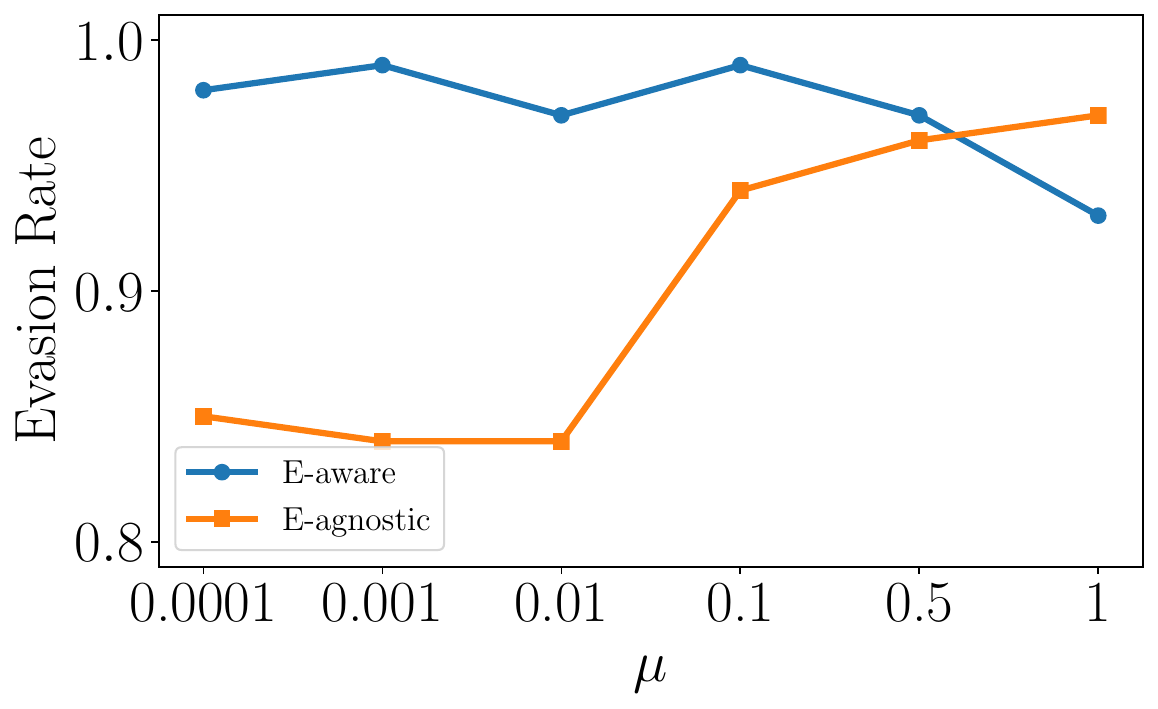}}
    \subfloat[Bitwise accuracy]{\includegraphics[width=0.33 \textwidth]{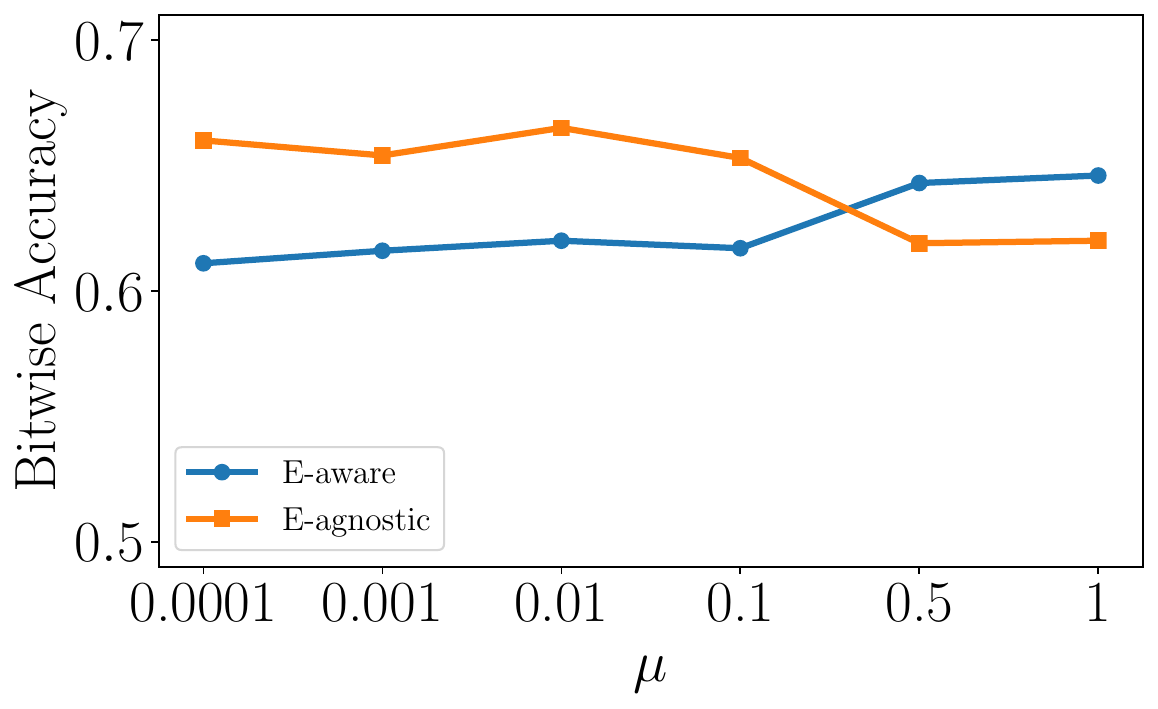}}
    \subfloat[FID]{\includegraphics[width=0.33 \textwidth]{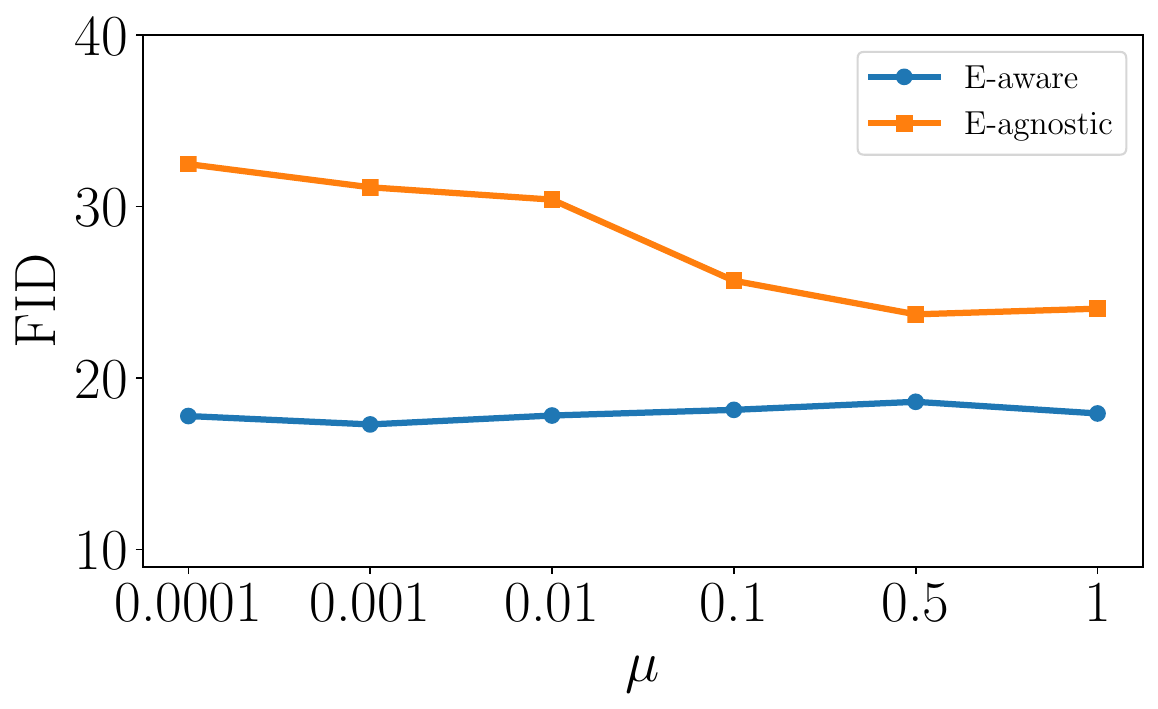}}
    \vspace{3mm}
    \caption{Effectiveness and utility of our attack with different $\mu$ on ImageNet.}
\label{fig-mu}
\end{figure}
Figure~\ref{fig-mu} shows the evasion rate, bitwise accuracy, and FID for different $\mu$ in our attack. First, we observe that the effectiveness of our attack remains constant, subsequently decreasing, while utility does not change in the E-aware scenario as $\mu$ increases. This phenomenon occurs because the reconstructed image is sufficiently similar to the non-watermarked image when $\mu$ is small; thus, increasing $\mu$ does not decrease their distance further to aid in watermark removal. Additionally, a larger weighting on perceptual loss reduces the mean square error's capability in removing watermark, resulting in decreased effectiveness of our attack. 
Conversely, in the E-agnostic scenario, both effectiveness and utility begin unchanged but later improve as $\mu$ increases. This improvement is attributed to the initial significant difference between the reconstructed and non-watermarked images. Increasing $\mu$ makes the reconstructed image closer to the non-watermarked one, thereby improving effectiveness and utility.
\section{Conclusion and Future Work}
In this work, we find that image watermark for open-source diffusion model is not robust as previously thought. Given a watermarked diffusion model, an attacker can remove the watermark from it by strategically fine-tuning its decoder. Our results show that our attack achieves both the effectiveness and utility goals in removing watermark from diffusion models in both E-aware and E-agnostic scenarios, and outperforms the existing model-targeted attack which is only applicable to E-aware scenario. Interesting future work is to design a more robust image watermarking method for open-source diffusion models.

\printbibliography

\newpage
\appendix

\begin{algorithm}  
\renewcommand{\algorithmicrequire}{\textbf{Input:}} 
\renewcommand{\algorithmicensure}{\textbf{Output:}}
    \caption{Estimate the denoised latent vector $z$}
    \label{algorithm:whitebox}
    \begin{algorithmic}[1]
        \REQUIRE Non-watermarked images $\{x^i\}_{i=1}^n$, watermarked decoder $D_w$, number of iteration for the first stage $n\_iter_1$, number of iteration for the second stage $n\_iter_2$, learning rate $\alpha$, perceptual loss function $l_p$
        \ENSURE Estimated denoised latent vectors $\{\hat{z}^i\}_{i=1}^n$
        \STATE $Q \gets \emptyset$
        \FOR{$i$ = 1 to $n$}
            \STATE $\hat{z}^i \sim \mathcal{N}(0, 1)$ 
            \FOR{$j$ = 1 to $n\_iter_1$}
            \STATE $g \gets \nabla_{\hat{z}^i}\|D_w(\hat{z}^i)-x^i\|_2$
            \STATE $\hat{z}^i \gets \hat{z}^i - \alpha \cdot g$
            \ENDFOR

            \FOR{$j$ = 1 to $n\_iter_2$}
            \STATE $g \gets \nabla_{\hat{z}^i}l_p(D_w(\hat{z}^i),x^i)$
            \STATE $\hat{z}^i \gets \hat{z}^i - \alpha \cdot g$
            \ENDFOR

            \STATE $Q \gets Q \cup \{\hat{z}^i\}$
            
        \ENDFOR
        \STATE return $Q$
    \end{algorithmic}
    \label{alg_1}
\end{algorithm}

\begin{algorithm}  
\renewcommand{\algorithmicrequire}{\textbf{Input:}} 
\renewcommand{\algorithmicensure}{\textbf{Output:}}
    \caption{Fine-tune the decoder $D_w$}
    \label{algorithm:whitebox}
    \begin{algorithmic}[1]
        \REQUIRE Non-watermarked images $\{x^i\}_{i=1}^n$, estimated denoised latent vectors $\{\hat{z}^i\}_{i=1}^n$, watermarked decoder $D_w$, number of epoch $n\_epoch$, decoder learning rate $\alpha$, discriminator learning rate $\beta$, perceptual loss function $l_p$, discriminator $disc$, weight for perceptual loss $\lambda$, weight for adversarial loss $\mu$
        \ENSURE Non-watermarked decoder $D_{nw}$
        \STATE $D_{nw} \gets D_w$
        \FOR{$i$ = 1 to $n\_epoch$}
            \STATE $g_{disc} \gets -\nabla_{disc} \frac{1}{n} \sum_{i=1}^{n} [log(1-disc(D_{nw}(\hat{z}^i))) + log(disc(x^i))]$
            \STATE $disc \gets disc - \beta \cdot g_{disc}$

            \STATE $g \gets \nabla_{D_{nw}}\frac{1}{n} \sum_{i=1}^{n} \| D_{nw}(\hat{z}^i) - x^i \|_{2} + \lambda \frac{1}{n} \sum_{i=1}^{n} l_p(D_{nw}(\hat{z}^i), x^i) + \mu \frac{1}{n} \sum_{i=1}^{n} log(1-disc(D_{nw}(\hat{z}^i)))$
            \STATE $D_{nw} \gets D_{nw} - \alpha \cdot g$            
        \ENDFOR
        \STATE return $D_{nw}$
    \end{algorithmic}
    \label{alg_2}
\end{algorithm}

\section{Details of the per-image-based removal attacks}
\label{apdx-per-image}
\begin{itemize}
    \item {\bf JPEG}
    It is a commonly used image compression technique that can significantly decrease the size of image files while preserving high image quality. The quality of images processed by JPEG is governed by a quality factor. Using a smaller quality factor to post-process watermarked images can make the detection of watermarks within the image more difficult.
    \item {\bf Brightness}
    This method modifies the brightness of an image by initially converting the image to a color space that includes a brightness-related channel. It then isolates this channel, adjusts its intensity by multiplying it with a specified factor, and finally converts the image back to its original color space. This method may disrupt the watermark patterns in watermarked images to evade watermark detection.
    \item {\bf Contrast}
    This method alters the contrast of an image by modifying its pixel values. Specifically, for each pixel, it subtracts 127 from the pixel's value, multiplies the result by a factor $k$, and then adds 127 to the outcome. The factor $k$ determines the level of contrast enhancement or reduction, with values greater than 1 increasing contrast and values between 0 and 1 decreasing it.
    \item {\bf Gaussian noise (GN)}
    This method adds a noise that follows a Gaussian distribution with a zero mean and a standard deviation of $\sigma$ to the watermarked image. It simulates the noise effects commonly encountered in the real world. A larger $\sigma$ value makes it more challenging to detect watermarks, simultaneously compromising image quality.
    \item {\bf WEvade-W-II~\cite{jiang2023evading}}
    This method employs projected gradient descent (PGD) to optimize a perturbation applied to the watermarked image such that the decoded watermark from the perturbed image by the model provider's watermarking decoder closely matches a randomly generated watermark, with each bit uniformly sampled from $\{0,1\}$. We assume that the attacker has access to the watermarking decoder for this method.
\end{itemize}

\section{Parameter settings for MP}
\label{apdx-mp}
Following the configuration by Fernandez et al.~\cite{fernandez2023stable}, we employ AdamW and a learning rate of 0.0005 with a linear warm-up period of 20 iterations followed by a half-cycle cosine decay to fine-tune the decoder with a batch size of 4 to achieve similar bitwise accuracy on the attacking dataset as our attack in the E-aware scenario.

\begin{figure}[t!]
\centering
\renewcommand*{\arraystretch}{0}
\begin{tabular}{*{5}{@{}c}@{}}
\begin{subfigure}{.19\linewidth}
  \centering
  \includegraphics[width=\linewidth]{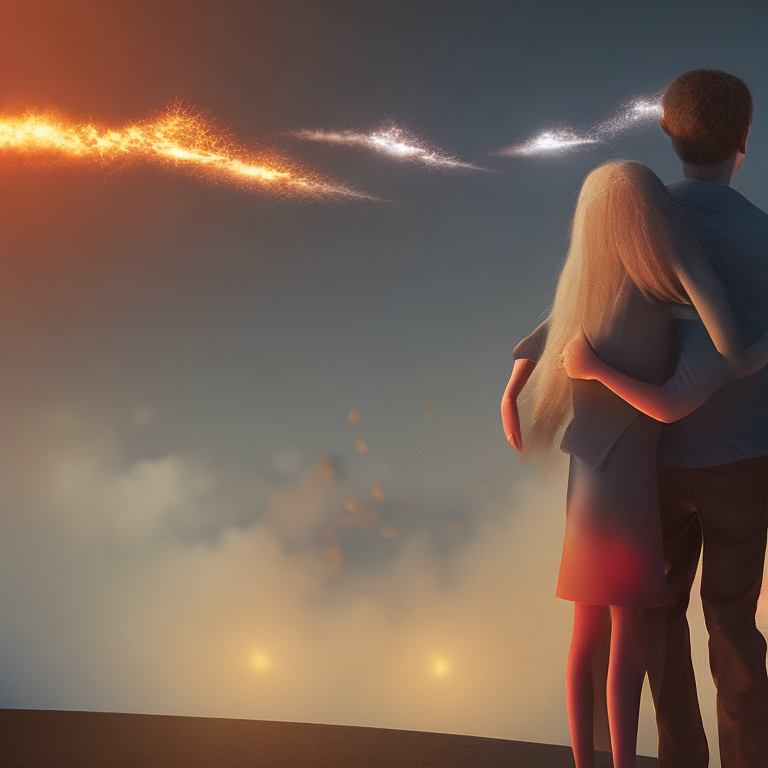}
\end{subfigure}
\begin{subfigure}{.19\linewidth}
  \centering
  \includegraphics[width=\linewidth]{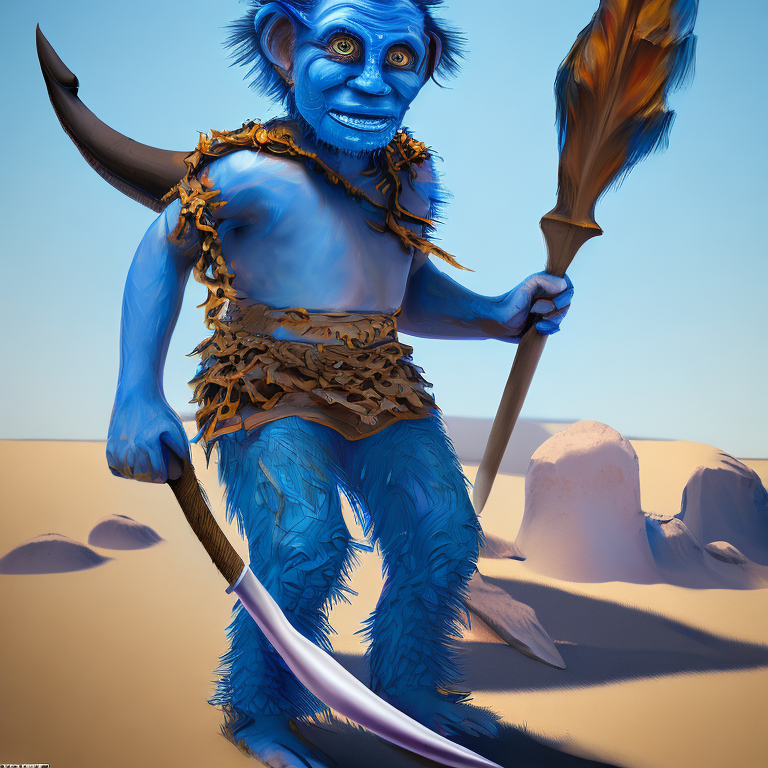}
\end{subfigure}
\begin{subfigure}{.19\linewidth}
  \centering
  \includegraphics[width=\linewidth]{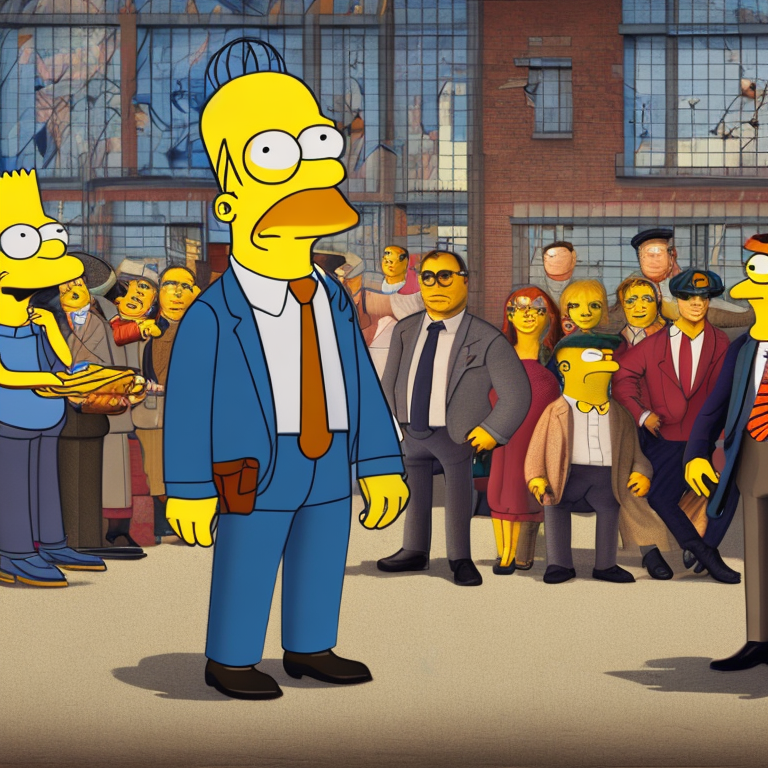}
\end{subfigure}
\begin{subfigure}{.19\linewidth}
  \centering
  \includegraphics[width=\linewidth]{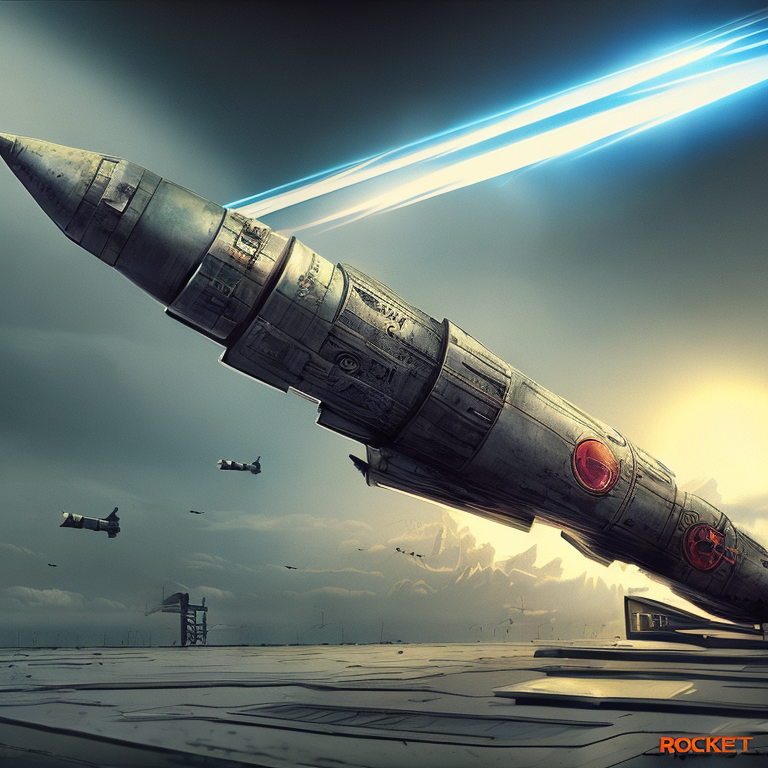}
\end{subfigure}
\begin{subfigure}{.19\linewidth}
  \centering
  \includegraphics[width=\linewidth]{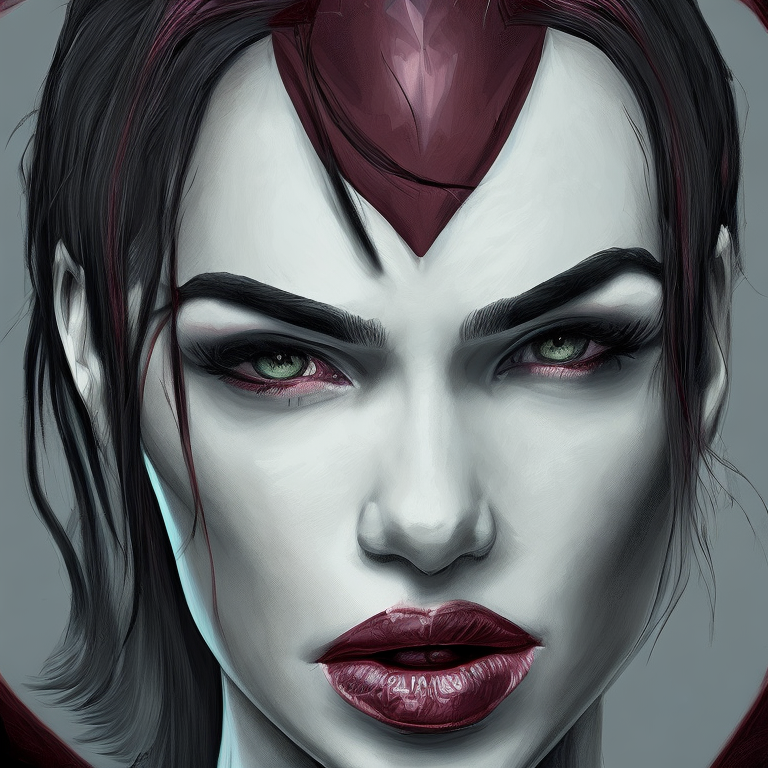}
\end{subfigure} \\

\begin{subfigure}{.19\linewidth}
  \centering
  \includegraphics[width=\linewidth]{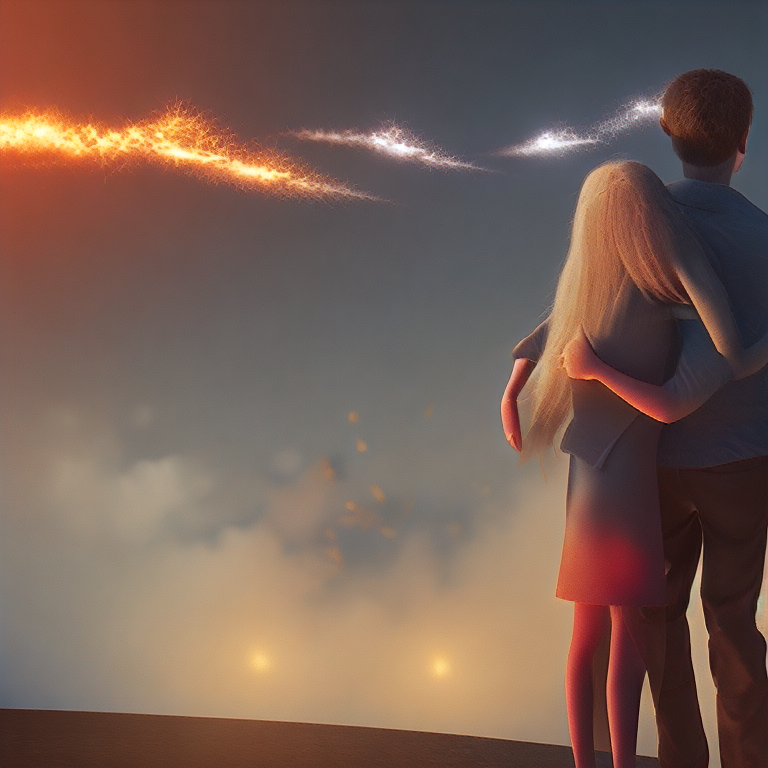}
\end{subfigure}
\begin{subfigure}{.19\linewidth}
  \centering
  \includegraphics[width=\linewidth]{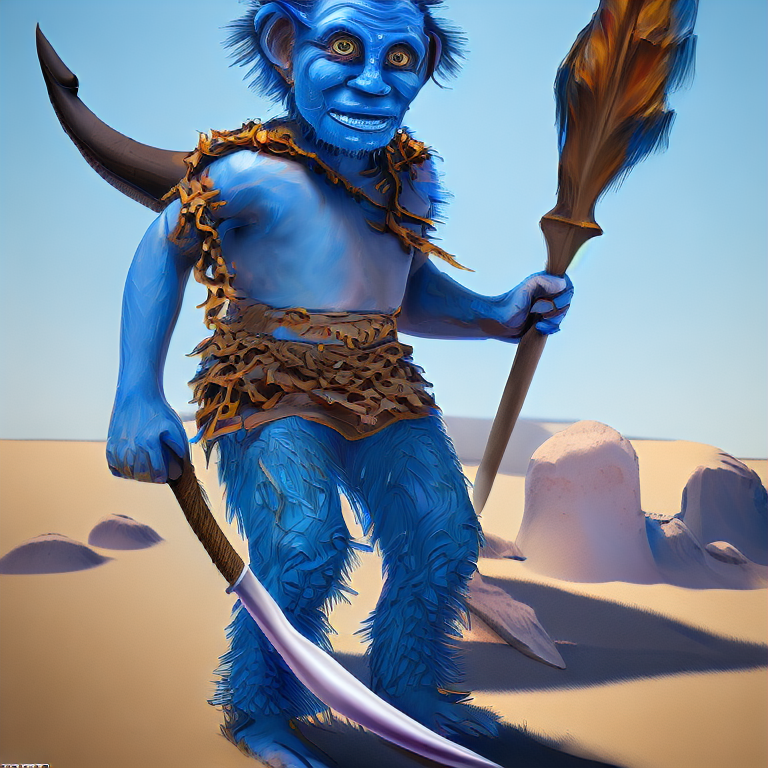}
\end{subfigure}
\begin{subfigure}{.19\linewidth}
  \centering
  \includegraphics[width=\linewidth]{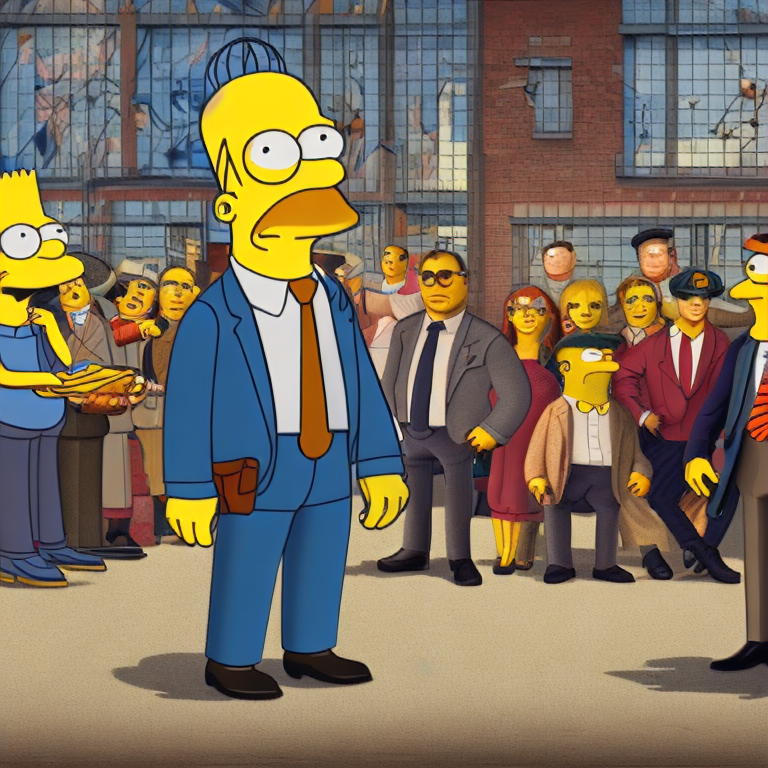}
\end{subfigure}
\begin{subfigure}{.19\linewidth}
  \centering
  \includegraphics[width=\linewidth]{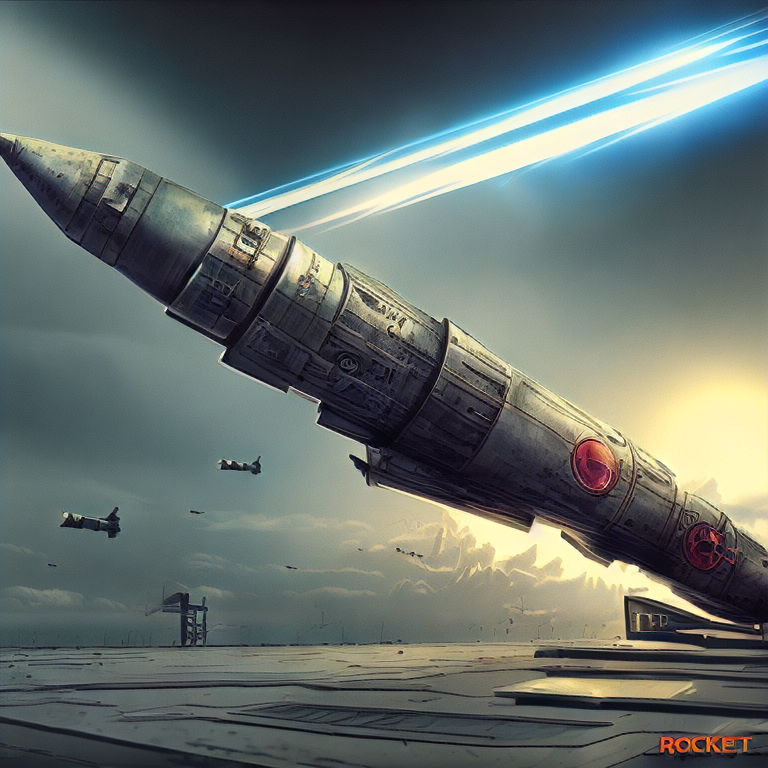}
\end{subfigure}
\begin{subfigure}{.19\linewidth}
  \centering
  \includegraphics[width=\linewidth]{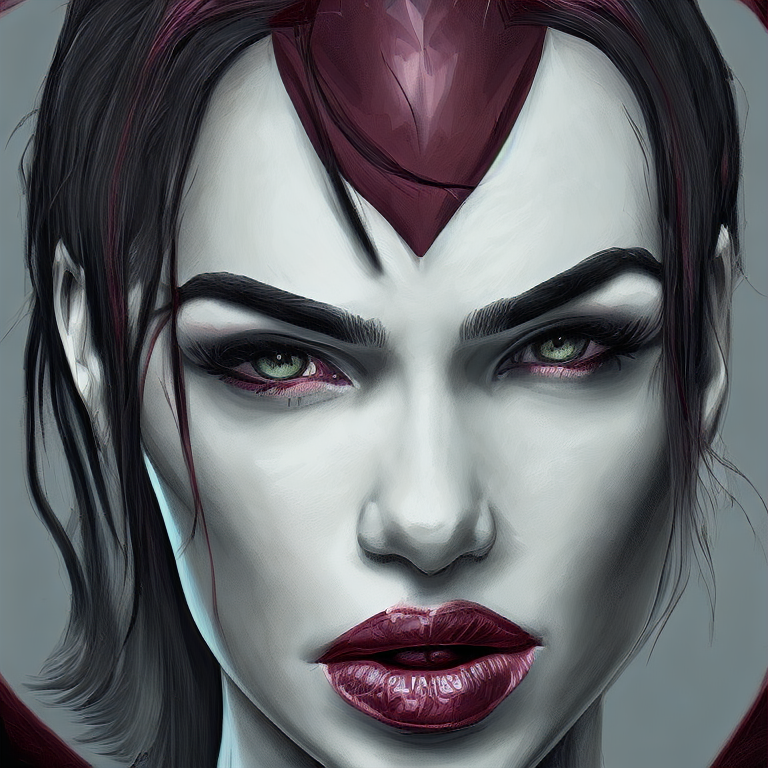}
\end{subfigure} \\

\begin{subfigure}{.19\linewidth}
  \centering
  \includegraphics[width=\linewidth]{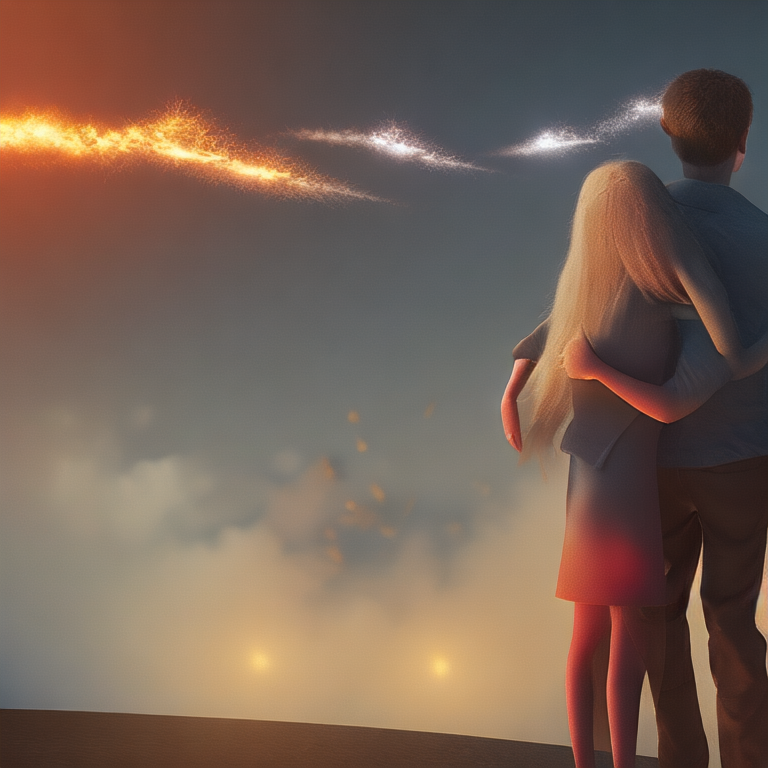}
\end{subfigure}
\begin{subfigure}{.19\linewidth}
  \centering
  \includegraphics[width=\linewidth]{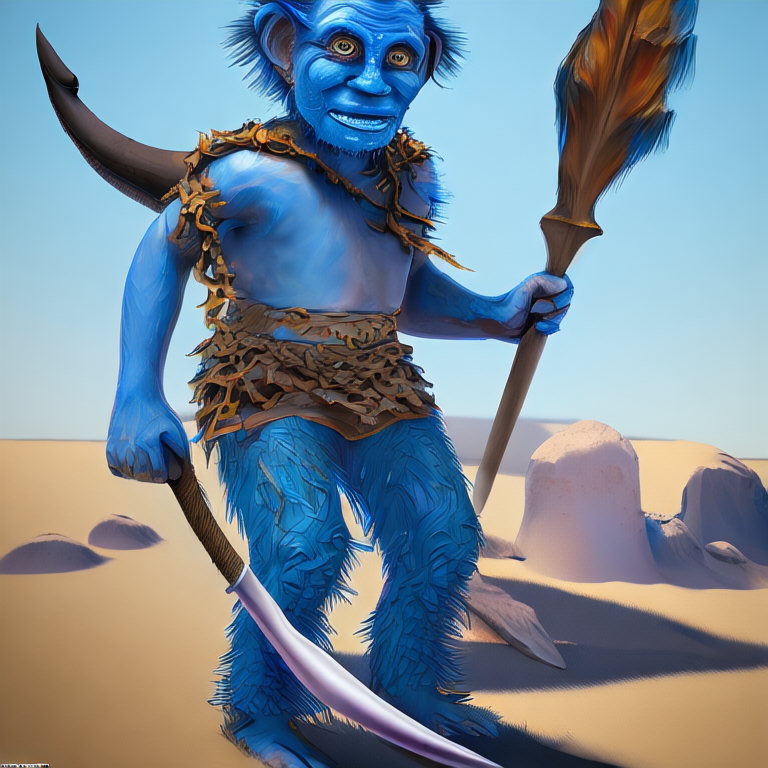}
\end{subfigure}
\begin{subfigure}{.19\linewidth}
  \centering
  \includegraphics[width=\linewidth]{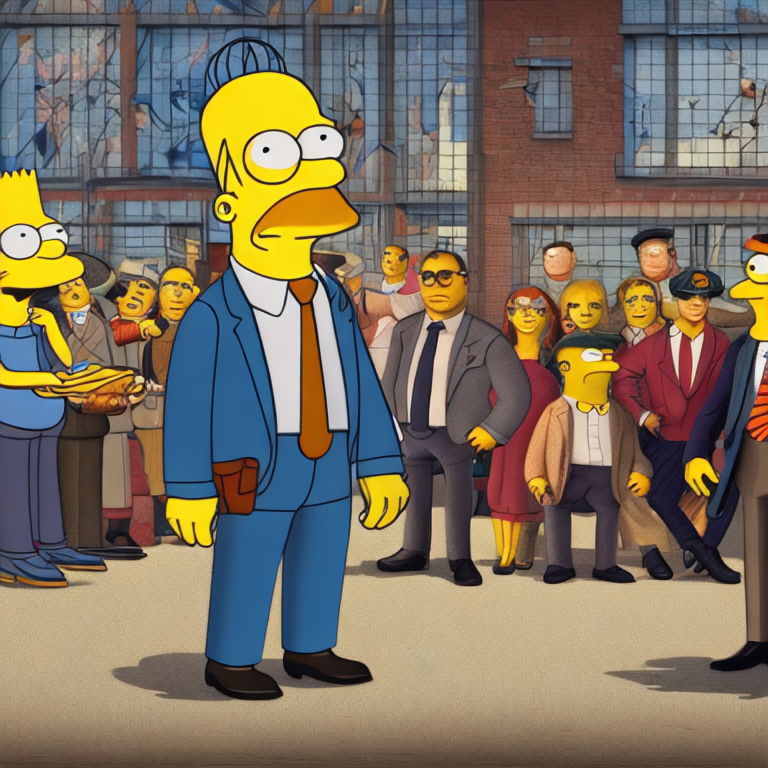}
\end{subfigure}
\begin{subfigure}{.19\linewidth}
  \centering
  \includegraphics[width=\linewidth]{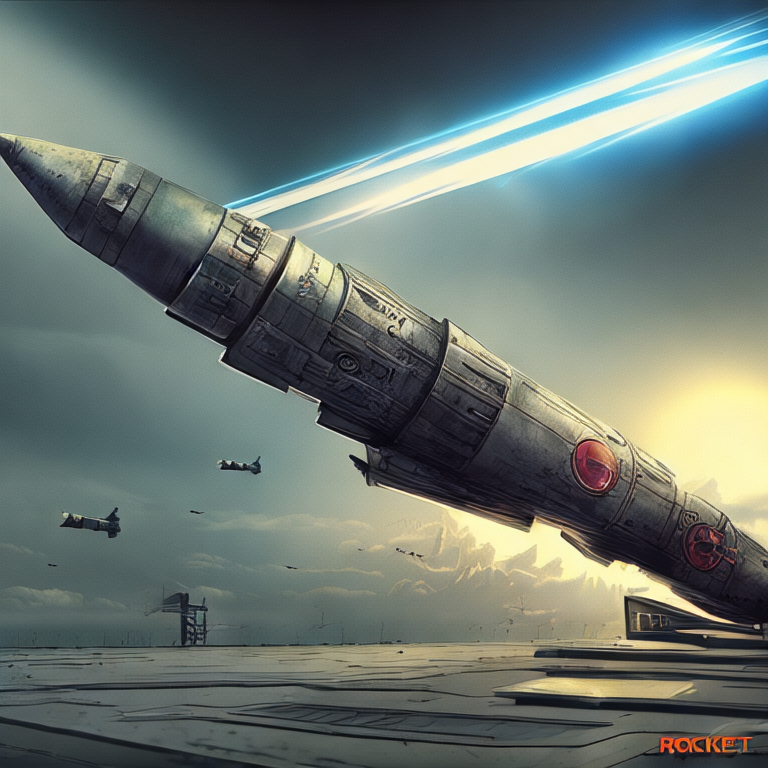}
\end{subfigure}
\begin{subfigure}{.19\linewidth}
  \centering
  \includegraphics[width=\linewidth]{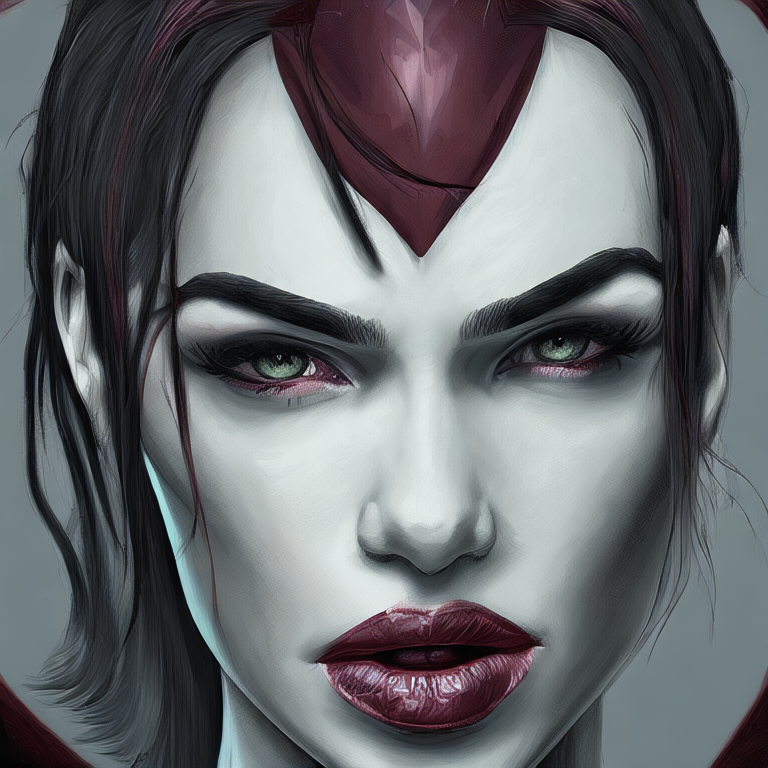}
\end{subfigure} \\

\begin{subfigure}{.19\linewidth}
  \centering
  \includegraphics[width=\linewidth]{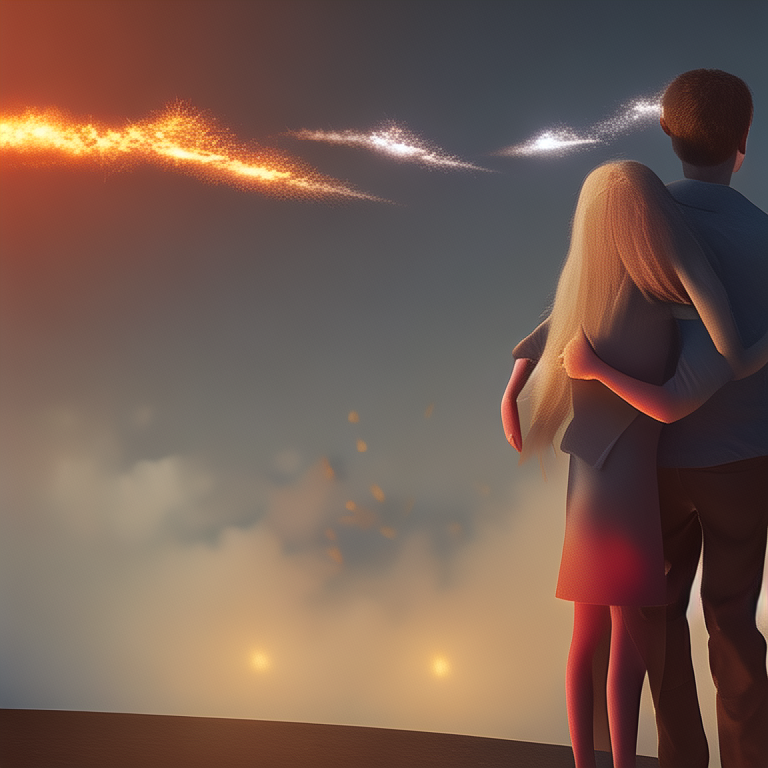}
\end{subfigure}
\begin{subfigure}{.19\linewidth}
  \centering
  \includegraphics[width=\linewidth]{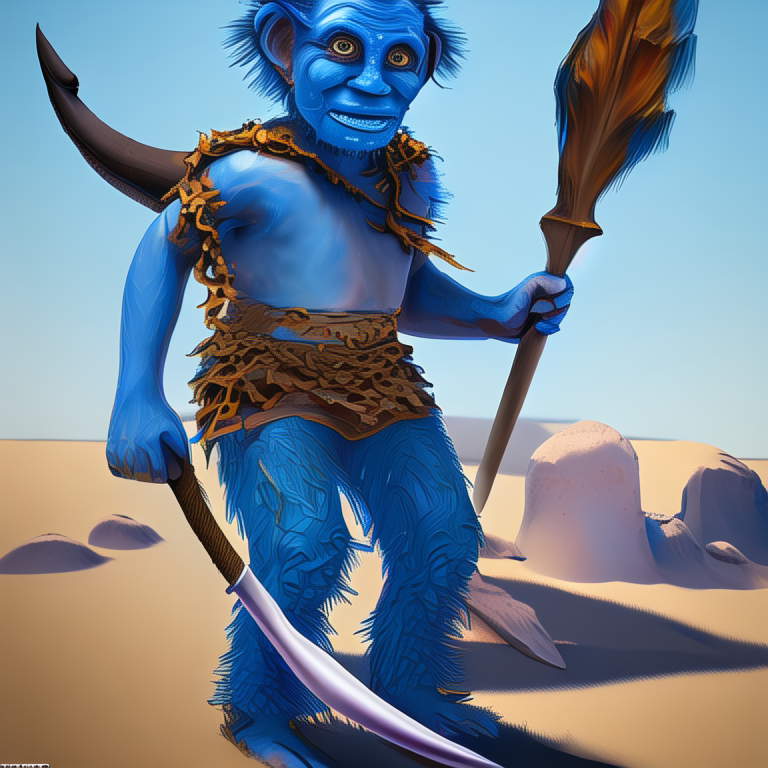}
\end{subfigure}
\begin{subfigure}{.19\linewidth}
  \centering
  \includegraphics[width=\linewidth]{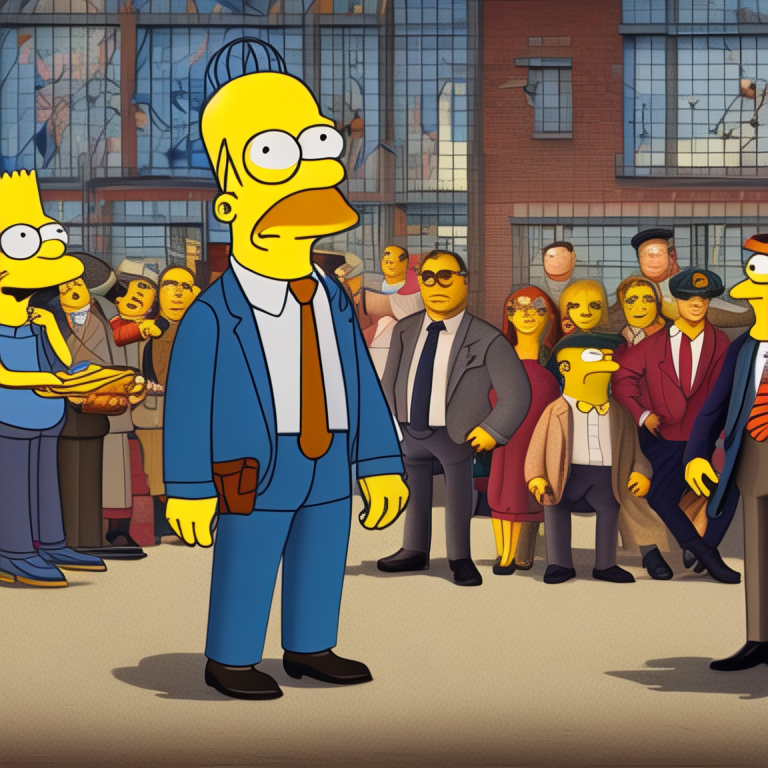}
\end{subfigure}
\begin{subfigure}{.19\linewidth}
  \centering
  \includegraphics[width=\linewidth]{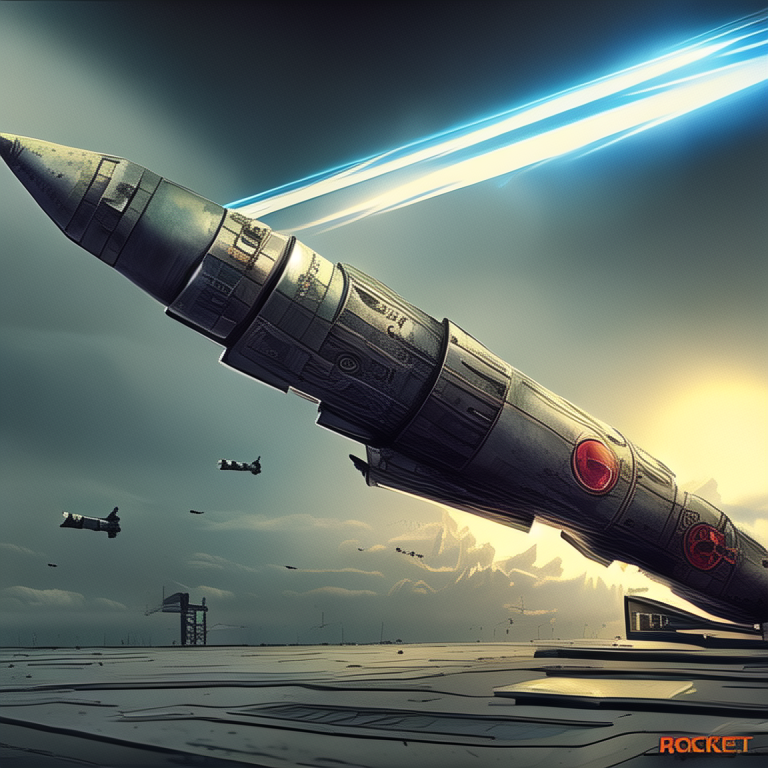}
\end{subfigure}
\begin{subfigure}{.19\linewidth}
  \centering
  \includegraphics[width=\linewidth]{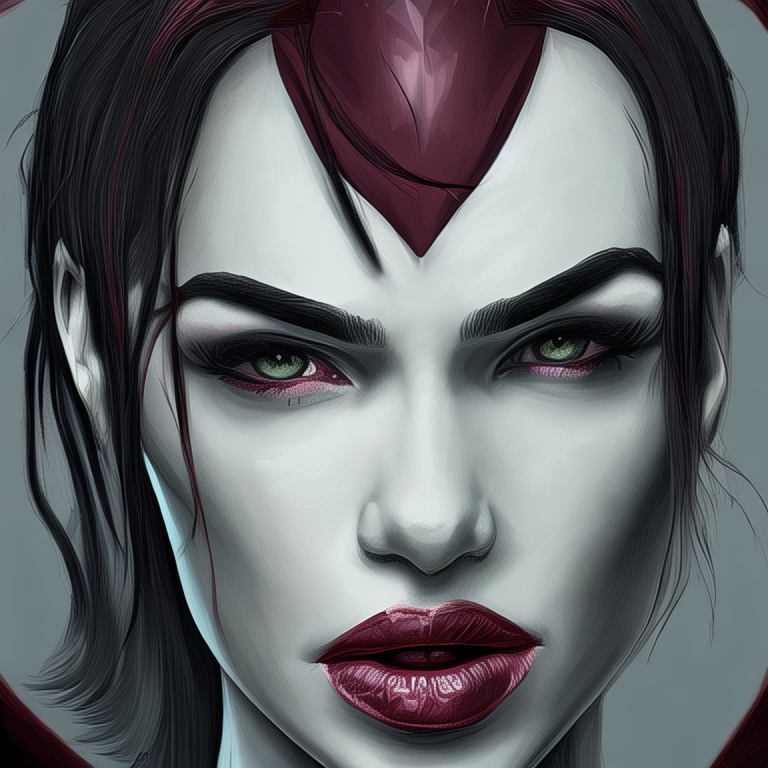}
\end{subfigure}
\end{tabular}

\vspace{3mm}
\caption{Image generated by the clean Stable Diffusion 2.1 (first row), Stable Diffusion 2.1 watermarked by Stable Signature (second row), watermarked Stable Diffusion 2.1 fine-tuned by our attack in E-aware scenario (third row), and watermarked Stable Diffusion 2.1 fine-tuned by our attack in E-agnostic scenario (fourth row). The same denoised latent vector is used by all diffusion models' decoders to generate the images in the same column. The watermark can only be detected in the images generated by Stable Diffusion 2.1 watermarked by Stable Signature (second row).}
\label{fig-comp}
\end{figure}

\begin{figure}[t!]
\centering
\begin{subfigure}{.19\linewidth}
  \centering
  \subfloat[Clean]{\includegraphics[width=\linewidth]{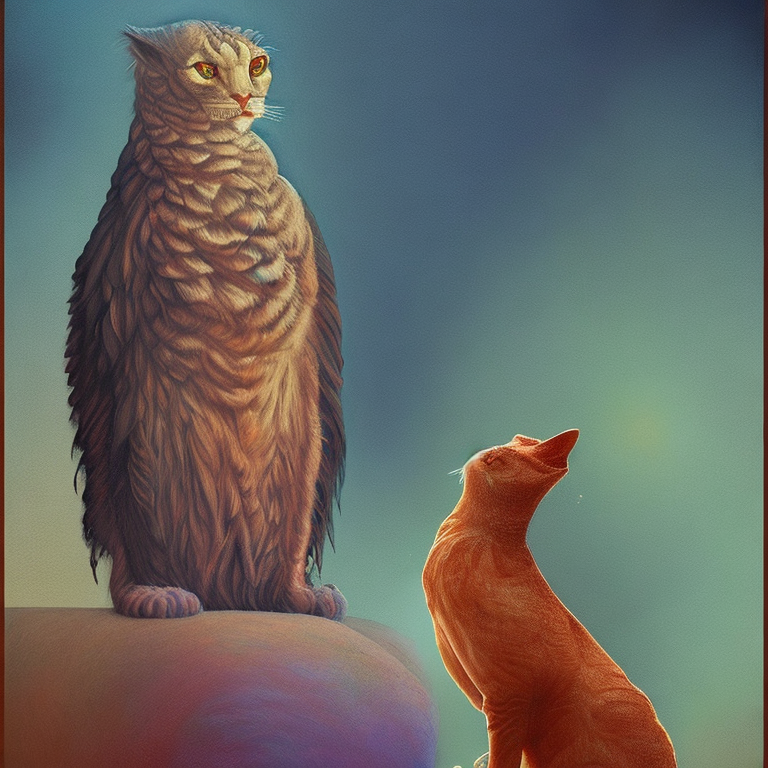}}
\end{subfigure}
\begin{subfigure}{.19\linewidth}
  \centering
  \subfloat[Watermarked]{\includegraphics[width=\linewidth]{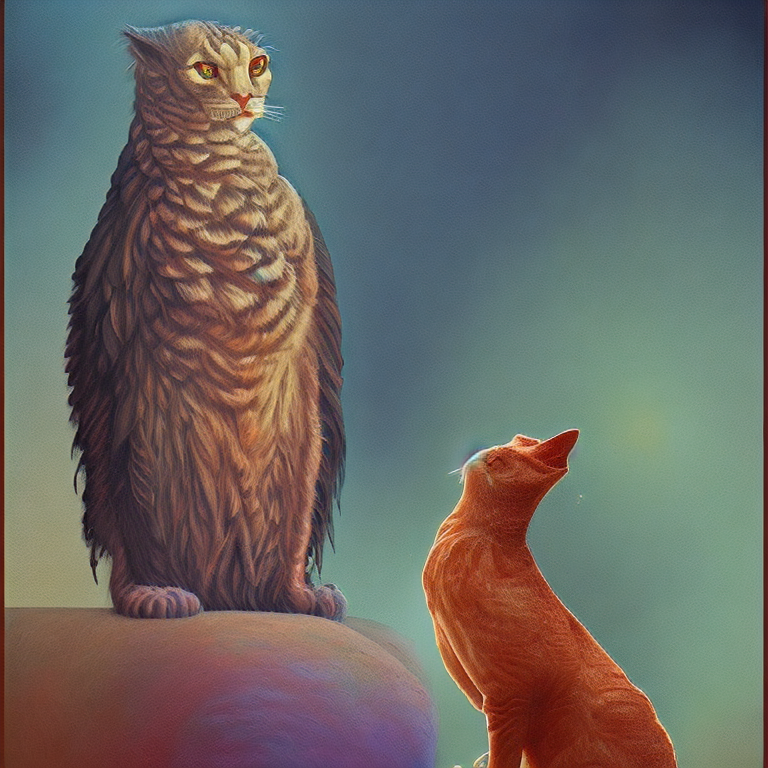}}
\end{subfigure}
\begin{subfigure}{.19\linewidth}
  \centering
  \subfloat[JPEG]{\includegraphics[width=\linewidth]{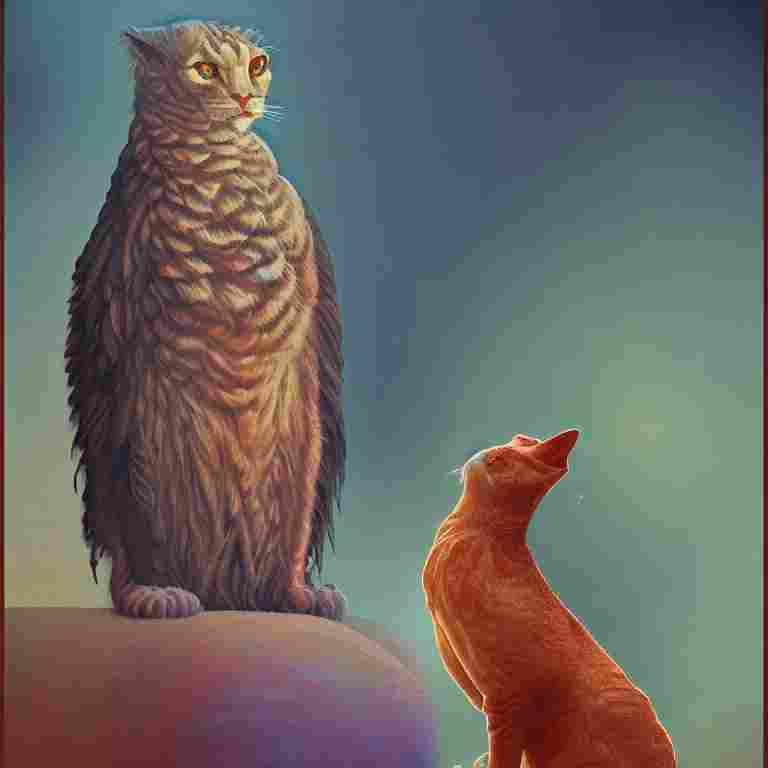}}
\end{subfigure}
\begin{subfigure}{.19\linewidth}
  \centering
  \subfloat[Brightness]{\includegraphics[width=\linewidth]{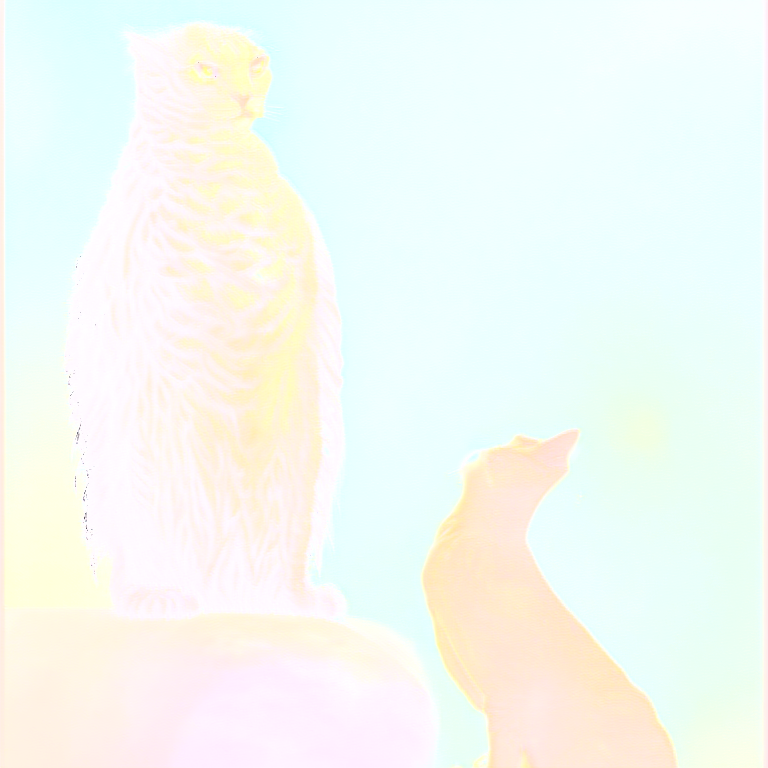}}
\end{subfigure}
\begin{subfigure}{.19\linewidth}
  \centering
  \subfloat[Contrast]{\includegraphics[width=\linewidth]{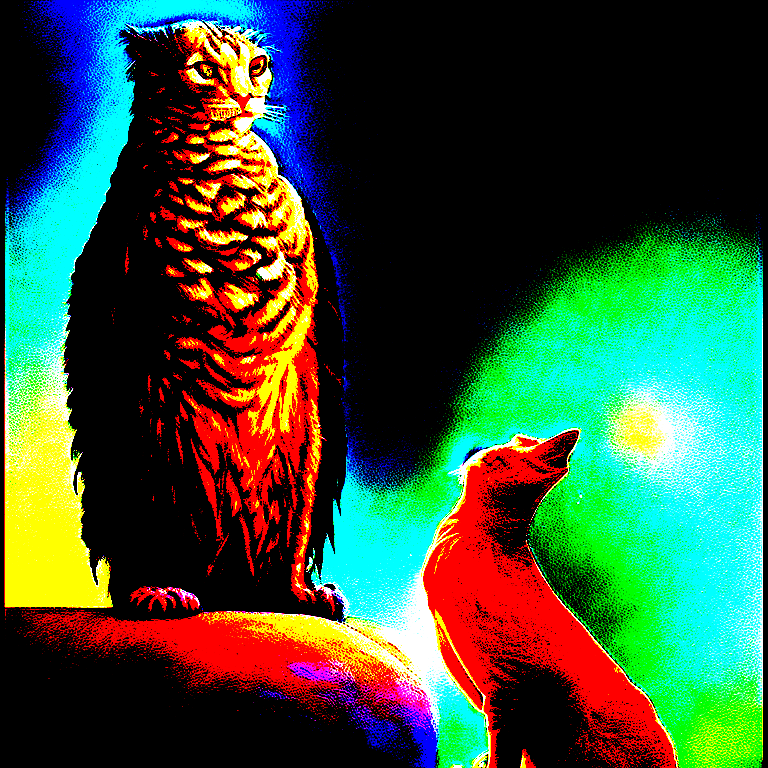}}
\end{subfigure} \\
\begin{subfigure}{.19\linewidth}
\centering
  \subfloat[GN]{\includegraphics[width=\linewidth]{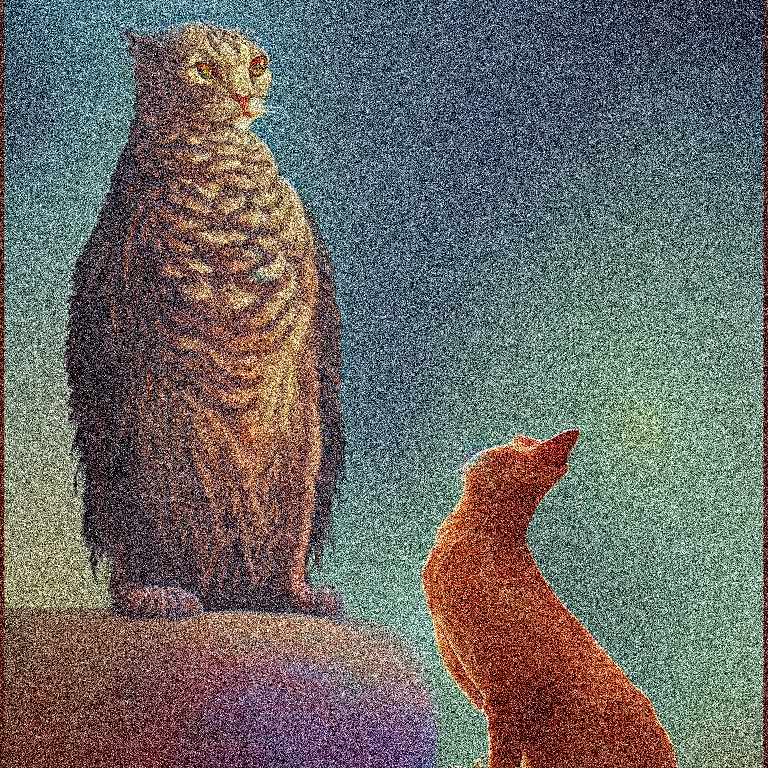}}
\end{subfigure}
\begin{subfigure}{.19\linewidth}
  \centering
  \subfloat[WEvade-W-II]{\includegraphics[width=\linewidth]{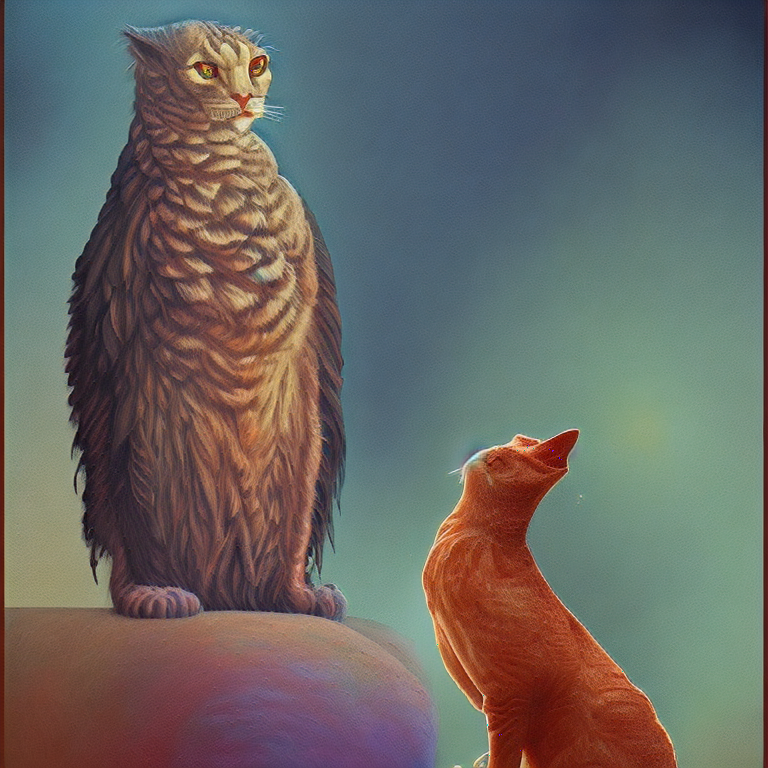}}
\end{subfigure}
\begin{subfigure}{.19\linewidth}
  \centering
  \subfloat[E-aware]{\includegraphics[width=\linewidth]{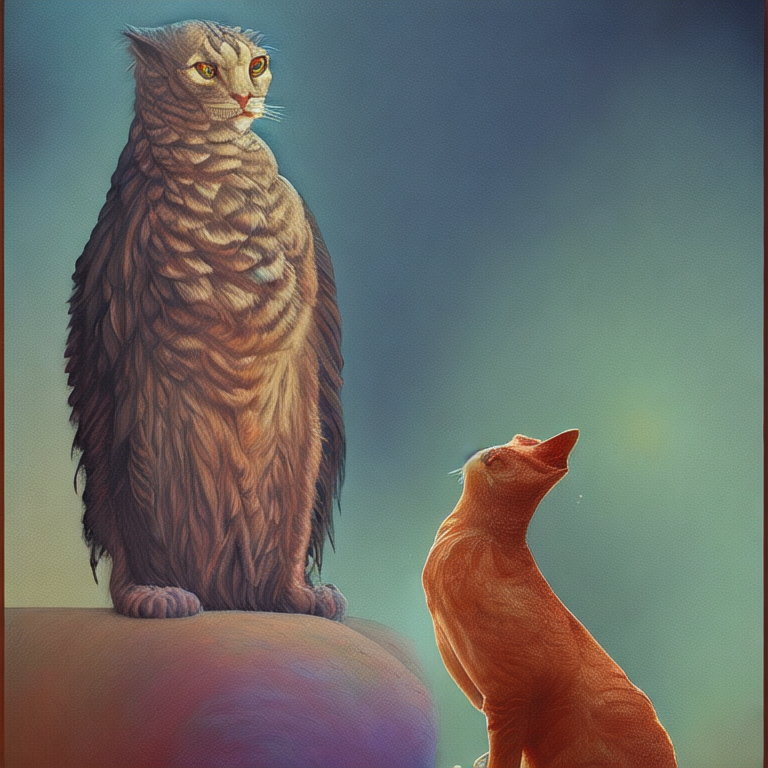}}
\end{subfigure}
\begin{subfigure}{.19\linewidth}
  \centering
  \subfloat[E-agnostic]{\includegraphics[width=\linewidth]{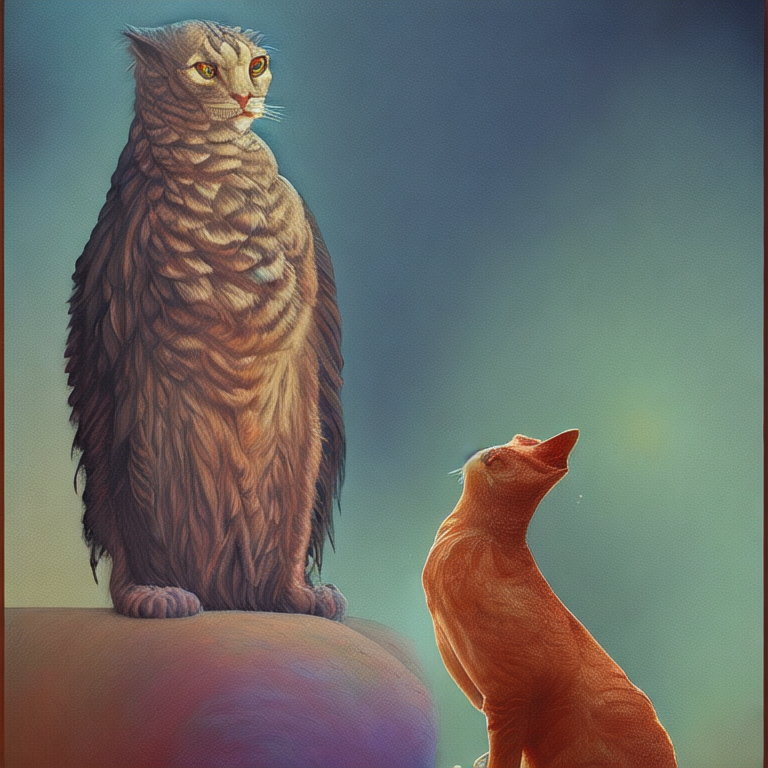}}
\end{subfigure}
\vspace{3mm}
\caption{An example of generated image (a) with clean decoder, (b) with watermarked decoder, (c) with watermarked decoder attacked by JPEG, (d) with watermarked decoder attacked by Brightness, (e) with watermarked decoder attacked by Contrast, (f) with watermarked decoder attacked by GN, (g) with watermarked decoder attacked by WEvade-W-II, (h) with non-watermarked decoder fine-tune by our attack in E-aware scenario, (i) with non-watermarked decoder fine-tune by our attack in E-agnostic scenario. The watermark can only be detected in (b).}
\label{fig-per-image}
\end{figure}

\end{document}